# BDoS: Blockchain Denial-of-Service

Michael Mirkin[*]
Technion and IC3
smirkin@campus.technion.ac.il

Yan Ji[*]
Cornell Tech and IC3
yj348@cornell.edu

Jonathan Pang
Cornell University
jp2268@cornell.edu

Ariah Klages-Mundt
Cornell University
aak228@cornell.edu

Ittay Eyal
Technion and IC3
ittay@technion.ac.il

Ari Juels
Cornell Tech and IC3
juels@cornell.edu

## ABSTRACT

Proof-of-work (PoW) cryptocurrency blockchains like Bitcoin secure vast amounts of money. Their operators, called *miners*, expend resources to generate blocks and receive monetary rewards for their effort. Blockchains are, in principle, attractive targets for Denial-of-Service (DoS) attacks: There is fierce competition among coins, as well as potential gains from short selling. Classical DoS attacks, however, typically target a few servers and cannot scale to systems with many nodes. There have been no successful DoS attacks to date against prominent cryptocurrencies.

We present *Blockchain DoS* (*BDoS*), the first *incentive-based DoS attack* that targets PoW cryptocurrencies. Unlike classical DoS, BDoS targets the system's *mechanism design*: It exploits the reward mechanism to discourage miner participation. Previous DoS attacks against PoW blockchains require an adversary's mining power to match that of all other miners. In contrast, BDoS can cause a blockchain to grind to a halt with significantly fewer resources, e.g., 21% as of March 2020 in Bitcoin, according to our empirical study. We find that Bitcoin's vulnerability to BDoS increases rapidly as the mining industry matures and profitability drops.

BDoS differs from known attacks like Selfish Mining in its aim not to increase an adversary's revenue, but to disrupt the system. Although it bears some algorithmic similarity to those attacks, it introduces a new adversarial model, goals, algorithm, and game-theoretic analysis. Beyond its direct implications for operational blockchains, BDoS introduces the novel idea that an adversary can manipulate miners' incentives by proving the existence of blocks without actually publishing them.

## 1 INTRODUCTION

Cryptocurrencies implemented with blockchain protocols based on Nakamoto's Bitcoin [88] have a current market capitalization of about $200B [29]. Like classical state machine replication protocols, blockchains allow participants to agree on a state—in their case, the client balances of a cryptocurrency. Unlike those classical protocols, however, public blockchains are decentralized and allow anyone to join the system at will.

To deter Sybil attacks [41], where an attacker masquerades as multiple entities, Nakamoto relies on *incentives*. Participants, called *miners*, expend resources and generate Proofs of Work (PoW) [42, 60]. They are rewarded with cryptocurrency for their efforts. Miners aggregate cryptocurrency *transactions* into so-called *blocks*, that also contain PoW. The blocks form a tree data structure, and a path in the tree is called a *blockchain*. The path representing the most work is called the *main chain*; its contents define the system's state.

An extensive line of work (§2) explores revenue-driven attacks against blockchains [45, 46, 68, 89, 102]. But DoS attacks, where the attacker is driven by exogenous incentives to stop a cryptocurrency blockchain, have received less attention. This may be because classical, network-based DoS attacks [40] do not scale to large decentralized systems; known mining-based DoS attacks [16, 17, 67] are prohibitively costly, as they require the attacker's mining resources to be at least equal to those of all other miners combined.

In this work, we present a new type of sabotage attack called *Blockchain Denial of Service* (*BDoS*). BDoS is incentive-based: The attacker targets the system's mechanism design and violates its incentive compatibility. Specifically, the attacker invests resources in order to incentivize rational miners to stop mining. A BDoS adversary can cause a blockchain to cease functioning with only a fraction of the resources of the other miners. It is the first incentive-based DoS attack of which we are aware.

The key element that enables BDoS is the consideration of miner behavior (§3) that was typically overlooked in previous work. First, miners can stop mining intermittently if it benefits them, as demonstrated empirically [28, 43, 69]; the majority of previous work assumes a static number of miners, i.e., miners always mine. Like prior work [23, 43, 116], we consider miners that do not venture on more elaborate strategic behavior [46, 89, 102], which indeed has not been observed in the wild.

Secondly, an attacker can signal to the miners that the system is in a state that reduces their revenue. Specifically, an attacker can generate a block and publish only a proof that she mined it, proving that she has an advantage over other miners, but without exposing the block's content. A natural candidate for such a proof is the *block header*. The profit of a rational miner decreases if she is not aware of the proof or ignores it, and therefore it is in her best interest to receive it (actively or passively). Similar behavior, where miners change their mining strategy as a result of external data, has been observed in practice [28, 43, 69]. Therefore, miners are expected to adapt in order to accept block header publication, even if currently this is not a standard publication method. Ignoring the block header is not an effective defense strategy as a miner is incentivized to defect from this countermeasure strategy to increase her payoff even at the expense of the common good. This situation is commonly referred to as *tragedy of the commons*, or *n-person prisoner's dilemma* [99].

The crux of the BDoS attack (§4) is as follows. The attacker generates a block $B_{\mathcal{A}}$ and publishes only its header (fig. 1b); we then say the attack is *active*. A miner can ignore the existence of the header of $B_\mathcal{A}$ and generate a block following its parent, resulting in a *fork* (fig. 1c). In this case, the attacker publishes the contents of $B_\mathcal{A}$, resulting in a *race* with two branches (fig. 1d). The miner's block might or might not end up in the main chain, depending on the parameters of the system. The implication is that the expected profitability of the rational miners decreases, and if it is low enough, then pausing mining becomes a better option than mining. If the profitability decrease is significant enough so that all miners stop mining, the attacker can cease mining as well, while she has an advantage of one block ($B_\mathcal{A}$). The blockchain thus grinds to a complete halt.

We formulate the behavior of the miners as a game and look for a dominant strategy (§5). The attack is successful when not mining is the best response of the miners. Success depends on several factors, mainly the sizes of the attacker and rational miners and the baseline profitability of mining. One might think that non-myopic miners invested in the success of the system would be willing to suffer a temporary profitability decline to overcome an attack and keep the blockchain running. However, we find that this is not so: If other miners behave altruistically and ignore the attack, a rational miner has an even stronger incentive to stop mining when the attack is active.

We consider several extensions of the action space. First, in practice, miners can mine on block headers, performing so-called *SPV Mining*. This action behavior is performed in practice by otherwise benign miners to slightly reduce latencies [97]. To overcome SPV mining we updated the attack (§6) as follows. If a rational miner successfully mines a block that extends the attacker's published header, the attacker abandons this header and never publishes its content, effectively invalidating the rational miner's block. We analyze the new game using *iterated elimination of strictly dominated strategies (IESDS)* [47], and show that stopping mining remains an equilibrium with the same parameters.

Secondly, we observe that the situation becomes significantly worse if miners have the option to use their resources in another blockchain rather than stop (§7). If two cryptocurrencies have similar baseline profitability, a small BDoS attacker can tip the scale and lead rational miners to defect from the attacked cryptocurrency to the now-more-profitable one.

Thirdly, we propose techniques for the attacker to prove she has a hidden block without exposing its header, making mitigation even harder (§8).

To empirically validate the practicality of BDoS, we calculate profitability in the longest-running cryptocurrency, Bitcoin. We combine mining difficulty data with mining hardware consumption and power, historical Bitcoin price fluctuation, and electricity costs. For example, as of March 13, 2020, given that the miners in Bitcoin have a $1.47 expected return on every $1 of electricity investment, an attacker with 21% of the mining power can successfully stop all rational miners. The instantaneous 50% drop in block reward (and thus profitability) that took place in 2020 [10] caused Bitcoin's security to be in even greater risk. Moreover, since the profitability of Bitcoin and Bitcoin Cash are almost identical [69, 113], the two-coin model implies that BDoS poses an imminent threat for both.

Constructively, we propose some possible mitigations to BDoS (§9). First, honest miners can prefer non-attacker blocks on a fork with a heuristic time-based detector. Secondly, alternative reward mechanisms [22, 123] compensate miners on lost races, making BDoS ineffective (though similar attacks might apply).

The discovery of BDoS adds another consideration for the evaluation of blockchain systems and raises questions on the existence of similar attacks against different blockchain designs (§10).

In summary:

- We introduce and explore new, practical actions in the action space of adversaries and miners.
- We initiate the first formal study of a mechanism-based DoS attack on PoW blockchains called Blockchain Denial-of-Service.
- We formalize a game between rational miners due to a BDoS adversary and show when the dominant strategy is to stop mining.
- We consider several extensions to the basic BDoS action / strategy space, including SPV mining, mining on other blockchains, and different methods of proving the existence of a mined block. We show that SPV mining doesn't help, and the other two strengthen the attack.
- We empirically study BDoS attacks in Bitcoin, showing that under reasonable assumptions, a BDoS attacker can succeed with roughly 21% mining power as of March 2020.
- We propose mitigations that can reduce the effectiveness of BDoS.

**Responsible disclosure** We have completed a disclosure process with prominent blockchain development groups.

## 2 RELATED WORK

To the best of our knowledge, this work is the first to study incentive-based denial of service attacks against blockchains. We present an overview here of previous work on denial-of-service attacks in the context of blockchains, incentive-related behavior, and other related work.

**DoS** Denial-of-Service (DoS) attacks [40] aim to prevent a system from serving clients, and are often mounted from multiple machines as *Distributed DoS* (*DDoS*) attacks. In blockchain networks, however, such techniques can only successfully target isolated system elements [61, 87, 118, 124] like cryptocurrency exchanges or mining coordinators in pools. In eclipse attacks [24, 106, 107] an adversary monopolizes all connections of a target node and isolates it from the network. When applied to blockchain systems [56, 76], the victim's local view is no longer in sync with the network, disrupting the victim and amplifying other blockchain attacks [89]. Similar effects can be achieved with routing attacks, chiefly BGP hijacking [4, 5, 114]. However, due to the decentralized structure of the system, nodes outside the effect of the attack can continue to interact with the blockchain as usual, apart from the possible reduction of attacked mining power. In contrast, BDoS stops all blockchain progress if other miners are rational.

Other attacks [72, 85, 86] saturate the blockchain to prevent transactions from being placed. Such attacks, however, result in graceful degradation, as the attacker simply raises the cost of transaction writes. Clients can still place transactions, albeit with a higher fee, thus also increasing the attacker's cost. Additionally, unlike BDoS, such attacks require continuous resource expenditure for the duration of the attack.

**Majority (51%) attacks** A 51% attack allows a miner that controls the majority of the mining power in the system to fork any section of the chain. She can mine on an old block and eventually build a longer chain than any minority competitors (even if the competitors have a significant head start). An attacker controlling a majority of the mining power violates the assumptions of PoW protocols and can perform a full-fledged DoS attack by simply generating empty blocks and ignoring other blocks. Since this is a majority attacker, her chain will extend faster than any other chain, making it the main chain, despite its empty content. An attacker with such power can also perform other attacks violating the system's safety properties. Goldfinger and bribery attacks [16, 17, 67, 73, 79, 115] utilize miner bribery to achieve similar effects, only without requiring the attacker to acquire mining power directly. Majority attacks have been observed in smaller cryptocurrencies [18, 38, 57], but not in major ones, possibly due to their high continuous cost. In contrast to this family of attacks, BDoS requires significantly lower than 50% mining-power budget, and no continuous expenditure.

**Revenue-seeking deviations** Nakamoto blockchains' security relies on an incentive mechanism that rewards miners that follow the rules. One line of study [8, 65, 77, 89, 91, 98, 102] considers incentive compatibility of blockchain protocols. It analyzes mining as a game, showing when the correct behavior is an equilibrium, and when deviations allow miners to increase their revenue. Such attacks may bias the mining power structure, leading to centralization, or affect other desired blockchain properties like censorship resistance. However, their goal and analysis consider only the internal system revenue, but not exogenous malicious motivations, and they cannot be directly applied to achieve denial of service.

Goren and Spiegelman [52] show that a miner can increase her revenue by mining intermittently. Unlike BDoS, this is a revenue seeking attack, only the attacker stops mining, and she is not manipulating the behavior of other miners.

Several incentive attacks can affect individual mining pools [45, 68, 71, 74, 100], but not directly lead to macro effects on blockchains.

**Incentive-based attacks** Another line of work explores attacks that use incentives to affect blockchain properties, using a form of bribery. Judmayer et al. [63] categorize incentives attacks by their goals into three groups: transaction revision, transaction ordering, and transaction exclusion. These attacks may not violate protocol safety directly, but can be used to force a particular order of transactions [33, 44, 103], or transaction omission [62, 79, 82, 122]. They do not affect the system liveness.

**Non-Nakamoto blockchains** The BDoS attack is designed specifically for Nakamoto-like blockchains. Nakamoto-like protocols with alternatives to PoW [25, 26, 120, 121, 125] are equally vulnerable. On the other hand, it does not directly apply to the Ethereum blockchain (that is more vulnerable to other attacks [91, 98], though), where blocks receive partial rewards even if they are off the main chain, so in case of a BDoS header publication, it's better for a participant to keep mining, getting at least a partial reward.

PoW alternatives such as Proof of Stake (PoS) [7, 34, 36, 51, 66] do not require participants to waste significant resources to approve transactions. Thus BDoS is not relevant to PoS in general. However, Buterin [21] introduced the so-called Discouragement Attack on PoS, where an attacker reduces the profit of other participants by censoring victims' messages, leading to a temporary DoS.

## 3 MODEL

We describe the system model (section 3.1), namely the participants, their interaction, and network assumptions, and the resultant *game model* (section 3.2), namely the miners' action space and utility function.

### 3.1 Mining Model

We model the system similarly to previous work [49, 83, 95] using common network assumptions [46, 89, 102]. However, we define an additional capability of the attacker. She can publish a proof that a block was mined, without publishing its content.

**Blockchain data structures** The system constructs a data structure called the *blockchain*, which is a a linked list of *blocks*. A block $B$ contains *block data* or payload, denoted by $D$, and metadata called *block header*, denoted by $H$. Thus, a block is a pair $B = (H, D)$. Each block header contains a hash reference to another block, except the so-called *genesis block* which we denote by $B_1$.

The blockchain is the main data structure in the system, and it defines the state of the cryptocurrency. Each block $B$ in the blockchain is either a *full block* containing the entire block information $(H, D)$, or a block header without the block data $(H, \perp)$ where $\perp$ denotes the lack of data. The fact that the blockchain can consist of partial block information is a refinement of our model compared to previous work [8, 46, 49, 83, 89, 91, 95, 98, 102], where a blockchain consists only of full blocks.

**Participants** We consider a system that comprises $n$ participants called miners, denoted by $\mathcal{P}_1, \mathcal{P}_2, \ldots, \mathcal{P}_n$, and an adversary $\mathcal{A}$. Each miner $\mathcal{P}_i$ has an associated value $\alpha_i$ called its *mining power*, and the adversary $\mathcal{A}$ has mining power $\alpha_{\mathcal{A}}$. The total mining power is normalized to 1, $\alpha_{\mathcal{A}} + \sum_{i=1}^{n} \alpha_i = 1$. Each miner has a *public key* known to all that allows her to prove her identity to other miners using a *private key*.

Each rational miner $\mathcal{P}_i$ possesses a view of the blockchain $L_i$ locally. $L_i^{Full}$ is the subset of $L_i$ that consists only of the full blocks in $L_i$ – i.e. blocks of the form $(H, D)$ where $D \neq \perp$.

$\mathcal{P}_i$ also has a local order function $O_i : L_i^{Full} \to \left\{1, \ldots, \left|L_i^{Full}\right|\right\}$. This function indicates the order of full blocks in $L_i$ observed by miner $\mathcal{P}_i$. Note that $O_i$ is not defined for blocks not in $L_i^{Full}$, i.e., partial blocks of the form $(H, \perp)$. For all $\mathcal{P}_i \in \{\mathcal{P}_1, \mathcal{P}_2, \ldots, \mathcal{P}_n\}$ it holds that $O_i(B_1) = 1$, that is, all miners agree that the genesis block is the first block. Different miners may have different order functions depending on the order they received their full blocks.

We call a path in the block tree consisting of *full* blocks a *chain*. The longest chain of full blocks in $L_i$ represents the state of the system from miner's $\mathcal{P}_i$ perspective and is called the *main chain*. If there are multiple longest chains, $\mathcal{P}_i$ considers the chain she observed first as the main chain, i.e., the chain whose last block $B$ has the smallest $O_i(B)$ value.

**Rushing** We denote by $\gamma$ the strength of $\mathcal{A}$'s rushing ability [46, 89, 102]. That is, if $\mathcal{A}$ publishes a block to compete with a newly published block by some other miner $\mathcal{P}_i$, $\gamma$ is the expected ratio of rational miners that adopt $\mathcal{A}$'s block. The remaining $(1-\gamma)$ are the miners that adopt $\mathcal{P}_i$'s block. Throughout the paper we consider mostly $\gamma = \frac{1}{2}$ in our examples, as, e.g., in [46, 102]. Ethereum developers even took measures to fix $\gamma = \frac{1}{2}$ [1], as they considered $\gamma > \frac{1}{2}$ to be achievable by potential selfish miners. Nevertheless, our results and most charts are generalized for any $0 \leq \gamma \leq 1$.

**Scheduler** The system progresses in *rounds*, orchestrated by a *scheduler*. On each round, the scheduler selects a miner to generate a new block. Messages are delivered immediately, and the system is synchronous. The pseudo-code of the scheduler is in Appendix E.

Each round has a duration. We denote by $\lambda$ a system constant called the *round rate constant*. It corresponds to the desired round rate (average number of rounds per second) in the blockchain. For instance, in Bitcoin $\lambda = \frac{1}{10 \cdot 60}$ s$^{-1}$, thus a block is created on average every 10 minutes.

In the beginning of each round $r$, the scheduler asks each miner whether she participates as a candidate to find a new block during this round. We say that a participating miner is *active* in this round. The scheduler also records the so-called *block template* of each active miner, which includes the miner's identity (using her public key) and the hash of the block it extends. Then the scheduler chooses a miner to mine the next block from the set of active miners by a weighted random distribution. Each miner's probability to be chosen is proportional to her mining power. The selected miner can create a block in round $r$ and is called the *winner* of the round, we denote it by $w_r$. We index the blocks $\mathbb{B} = \{B_1, B_2, \ldots\}$ by the order of their issuance, i.e., $w_r$ creates block $B_r$. We denote by $\alpha_{\text{active}}^r$ the total mining power of active miners in round $r$. The scheduler determines the duration of the round, using an exponential distribution with rate $\lambda \cdot \alpha_{\text{active}}^r$. Note that we do not consider difficulty adjustment unless otherwise stated; thus, the expected block generation time in a round is $\frac{1}{\lambda \cdot \alpha_{\text{active}}^r}$. If all the miners are mining during a round $r$ (i.e., $\alpha_{\text{active}}^r = 1$), which we call the *honest setting*, it holds that the exponential distribution of the duration of round $r$ has a rate of $\lambda$.

Next, the scheduler adds the partial or full block to the private ledgers of all miners. It treats the cases of an adversarial winner and a rational winner separately. If the adversary $\mathcal{A}$ wins, she decides whether to publish the full block of $B_r$ or only the block header. She then announces her decision to the scheduler. Receiving the adversary's decisions, the scheduler adds to the private ledgers of the other miners either the full block or the block header of $B_r$.

If a rational miner $\mathcal{P}_i$ wins, the scheduler first notifies the adversary $\mathcal{A}$ of $B_r$. The adversary decides whether to race against $B_r$. If she decides to race, she sends the full block that corresponds to a previously withheld block. Otherwise, the adversary sends an empty message. If the message is empty, the scheduler simply broadcasts $B_r$ to all miners. Otherwise, the scheduler sends $B_r$ and $\mathcal{A}$'s competing blocks in different orders to different miners, to simulate the connectivity factor $\gamma$: For each miner $p \in \{\mathcal{P}_1, \ldots, \mathcal{P}_n\} \setminus \{w_r\}$, with probability $\frac{\gamma(1-\alpha_\mathcal{A})}{1-\alpha_\mathcal{A}-\alpha_{w_r}}$ the scheduler sends $\mathcal{A}$'s competing blocks first and then $B_r$ to $p$, and with probability $1 - \frac{\gamma(1-\alpha_\mathcal{A})}{1-\alpha_\mathcal{A}-\alpha_{w_r}}$ it sends $B_r$ first and then $\mathcal{A}$'s blocks.

## 3.2 Game-Theoretic Model

The system model gives rise to a game played among the rational miners given the adversary's behavior.

**Players** The players are the miners where each rational miner $\mathcal{P}_i$ possess a mining power $\alpha_i$. Each miner knows the adversary's attack protocol.

**Utility** For each miner $\mathcal{P}_i$ we denote by $\Pi_i(t)$, $R_i(t)$, and $C_i(t)$ her expected profit, revenue, and cost until time $t$, respectively. It holds that: $\Pi_i(t) = R_i(t) - C_i(t)$. We denote the average revenue and cost per time unit, for $\mathcal{P}_i$ by $\hat{R}_i \triangleq \lim_{t \to \infty} \frac{R_i(t)}{t}$ and $\hat{C}_i \triangleq \lim_{t \to \infty} \frac{C_i(t)}{t}$ respectively. Consequently, the average profit per time unit, for $\mathcal{P}_i$, is: $\hat{\Pi}_i \triangleq \hat{R}_i - \hat{C}_i$. Note that any one-time cost is negligible when we discuss an infinite-horizon game.

For simplicity, we assume that the coin's real value is constant during the entire game, and denote the per-block mining reward by $K$. We ignore the effect of varying transaction fees, since the transaction fee is typically negligible [14] and does not change the results significantly. Different miners may mine at different electricity costs, but we assume that the cost does not change throughout the game. Thus, we do not consider the difficulty adjustment. This assumption is reasonable in Bitcoin and other cryptocurrencies where the difficulty adjustment happens infrequently. Therefore, $\mathcal{A}$ can choose the timing of the attack to be long before the next difficulty adjustment.

The cost of miner $\mathcal{P}_i$ per one second of mining is $\alpha_i c_i$, where $c_i$ is the normalized mining cost per second for $\mathcal{P}_i$. Therefore, when there is no attack, the expected profit per time unit is $\hat{\Pi}_i^b = \alpha_i(\lambda K - c_i)$.

In order to define the utility function, we normalize the expected profit by the miner's mining power. The utility function $U$ of $\mathcal{P}_i$ is thus: $U_i \triangleq \frac{\hat{\Pi}_i}{\alpha_i}$.

We conclude that the utility of the rational miner $\mathcal{P}_i$ during an honest game (with no attack) is:

$$U_i^b \triangleq \lambda K - c_i. \tag{1}$$

We also define the *profitability factor* $\omega_i^b$ for miner $\mathcal{P}_i$ participating in an honest game. The profitability factor is the return per dollar investment for a miner in an honest game:

$$\omega_i^b \triangleq \lim_{t \to \infty} \frac{R_i(t)}{C_i(t)} = \frac{\lambda K}{c_i}. \tag{2}$$

We note that $U_i^b > 0$ implies $\omega_i^b > 1$ and $U_i^b < 0$ implies $\omega_i^b < 1$.

**Actions** We consider miners that are rational, meaning that they do not participate in the game when it is not profitable. The miners are trying to maximize their profit within the protocol rules, with the ability to exit the game – i.e. stop mining. Specifically, each rational miner has two possible actions:

(1) mine - Mine on the main chain, or
(2) stop - Stop mining.

A miner chooses an action at the beginning of a round by notifying the scheduler. Note that switching the action within the round does not make sense, since no new information is available to $\mathcal{P}_i$ during a round. The elapsed time does not provide any new information due to the memorylessness property [102, 116]. This is formally justified in Appendix A.

Note that if there is no attack and $\omega_i^b > 1$, the rational miner always chooses mine and if $\omega_i^b < 1$ she chooses stop.

In case the adversary releases a block header, a rational miner has to choose one of the two actions. The pseudocode that describes the rational miner's possible actions is in Appendix E.

## 4 THE BDOS ATTACK

The BDoS attack aims to incentivize rational miners to stop mining. The crux is that an attacker ($\mathcal{A}$) can bring the system to a state where if a rational miner $\mathcal{P}_i$ chooses mine and finds a block $B_{\mathcal{P}_i}$, $\mathcal{A}$ can invalidate $B_{\mathcal{P}_i}$—with some probability. Thus, while $\mathcal{P}_i$ incurs the same cost for performing mining (including electricity and other varying costs) as in the honest game, there is a smaller probability that $\mathcal{P}_i$ would see any reward for her investment.

We now describe the strategy, which is illustrated in fig. 1. Denote by $B^*$ the latest block on the main chain. $\mathcal{A}$'s attack algorithm is to mine on $B^*$ (fig. 1a). If she successfully appends a new block $B_{\mathcal{A}} = (H_{\mathcal{A}}, D_{\mathcal{A}})$ to $B^*$, rather than publishing $B_{\mathcal{A}}$ in full, she publishes *only its header* $(H_{\mathcal{A}}, \bot)$. She withholds the rest of the block, namely its associated transactions. At this point, we refer to the state of the attack as *active*. We refer to $B_{\mathcal{A}}$ as the *leading block*. $B_{\mathcal{A}}$ is not part of the main chain, as it has not been published in full (fig. 1b).

The header of $B_{\mathcal{A}}$ serves as a proof that $\mathcal{A}$ has successfully mined $B_{\mathcal{A}}$ and is currently withholding the full block. Until another miner produces a new block, $\mathcal{A}$ stops mining completely. If all other miners also stop, the system thus reaches a standstill. But if at least one miner performs mine it would eventually generate a block $B_{\mathcal{P}_i}$ appended to $B^*$ (fig. 1c). In this case $\mathcal{A}$ immediately publishes $B_{\mathcal{A}}$ in full, i.e., attempts to add it to the main chain. A race ensues as describe in §3: Mining power is now divided between $B_{\mathcal{P}_i}$ and $B_{\mathcal{A}}$ (fig. 1d). The first block to be extended "wins" the race and becomes part of the main chain.

The effect of the attack on $\mathcal{P}_i$'s actions depends on the values of the system parameters $\omega_i^b$, $\alpha_i$ and $\alpha_{\mathcal{A}}$. The pseudocode for *BDoS* is in Appendix E.

## 5 ANALYSIS

If stop is the best response for all miners, we say the attack is successful as it achieves a *complete shutdown* of the system. If stop is best for some miners, we say BDoS is *partially successful* as it only slows the blockchain's progress, leading a *partial shutdown*.

### 5.1 Game-Theoretic Analysis

We now derive the possible strategy space for a rational miner. We analyze the game as an infinite-horizon game where the miners play indefinitely [46, 110]. This applies although the cryptography in the Nakamoto consensus breaks in an infinite game – as we analyze an ergodic process, the average utility over infinite time is similar to the average utility of finite games. Therefore, we are interested in the expected profit per second of the miners that would allow us to compare different strategies. In order to calculate it, we construct a Continuous-Time Markov Chain for each strategy. Unlike previous analysis of similar games [46], the block creation rate varies when the attack is active/inactive, and therefore our system cannot be described with a discrete-time Markov chain. The Markov chains allow us to compute the utility function for each strategy as a function of the other players' strategies. We analyze the conditions for a specific strategy (that corresponds to stop mining) to be the *dominant strategy* by comparing its utility to other strategies given the same choice of the other players. The analysis in this section assumes that $\omega_i^b > 1$ for all $\mathcal{P}_i$, i.e., honest mining is profitable for all players in the absence of an attack.

**Strategies** We evaluate the strategies from the perspective of a rational miner $\mathcal{P}_i$. We denote by $\Lambda_{B^*}$ the set of miners actively mining on $B^*$ while the attack is active. Next, we define $\alpha_{B^*}$ to be the hash rate of all miners in $\Lambda_{B^*}$. Given the attack algorithm BDoS and honest game profitability $\omega_i^b$, our goal is to find an optimal strategy for $\mathcal{P}_i$ which she chooses at the beginning of the game, i.e., a map from the private ledger $L_i$ and the order function $O_i$ to an optimal action. We say that strategy $S_1$ is more *beneficial* than strategy $S_2$, for a rational $\mathcal{P}_i$, if the utility by playing $S_1$ is larger than the utility by playing $S_2$. We assume that $\omega_i^b > 1$ (as for $\omega_i^b \le 1$ no rational miner mines). Therefore, a rational miner always chooses action mine when the attack is not active. We prove this intuitive claim in Appendix B. Consequently, we consider only two strategies, $S_{\text{mine}}$ and $S_{\text{stop}}$, that differ by the actions of $\mathcal{P}_i$ during the attack: mine and stop, respectively. We describe the game for each strategy with a three-state Markov chain. Strategy $S_{\text{mine}}$ appears in fig. 2a and $S_{\text{stop}}$ in fig. 2b. In both chains, state 0 represents the initial state where everyone mines on $B^*$ (fig. 1a). State 1 represents the state where the adversary managed to find a block (fig. 1b). State 2 represents the race condition, where the miners are divided between $\mathcal{A}$'s block and the block generated by a rational miner (fig. 1c). In both strategies, $\mathcal{P}_i$ chooses the action mine when not in state 1 (when the attack is not active).

**State Probabilities** We denote $\mathcal{P}_i$'s strategy by $S$. We can express $\alpha_{B^*}$ as a function of $S$:

$$\alpha_{B^*}(S) \triangleq \begin{cases} \alpha_{B^*} + \alpha_i, & \text{if } S = S_{\text{mine}} \\ \alpha_{B^*}, & \text{otherwise.} \end{cases} \quad (3)$$

We proceed to calculating the state probabilities of the two Markov chains in fig. 2:

$$p_0^S = \frac{\alpha_{B^*}(S)}{\alpha_{\mathcal{A}} \cdot \alpha_{B^*}(S) + \alpha_{\mathcal{A}} + \alpha_{B^*}(S)},$$
$$p_1^S = \frac{\alpha_{\mathcal{A}}}{\alpha_{\mathcal{A}} \cdot \alpha_{B^*}(S) + \alpha_{\mathcal{A}} + \alpha_{B^*}(S)}, \quad (4)$$
$$p_2^S = \frac{\alpha_{\mathcal{A}} \cdot \alpha_{B^*}(S)}{\alpha_{\mathcal{A}} \cdot \alpha_{B^*}(S) + \alpha_{\mathcal{A}} + \alpha_{B^*}(S)}.$$

Note that miner $\mathcal{P}_i$ changes the state probabilities depending on which strategy she chooses, as $\alpha_{B^*}(S)$ depends on $\mathcal{P}_i$'s strategy.

**Utility For Each Strategy** As the first step in calculating the utility, we calculate the cost and the revenue of $\mathcal{P}_i$. While a rational miner is mining, her cost per second is constant. However, when she stops mining, her cost per second is zero. Therefore for $S_{\text{stop}}$ it holds that the average cost per time unit $\hat{C}_i^{S_{\text{stop}}}$ for $\mathcal{P}_i$ is:

$$\hat{C}_i^{S_{\text{stop}}} = \lim_{t \to \infty} \frac{C_i^{S_{\text{stop}}}(t)}{t} = \alpha_i (1 - p_1^{S_{\text{stop}}}) \cdot c_i.$$

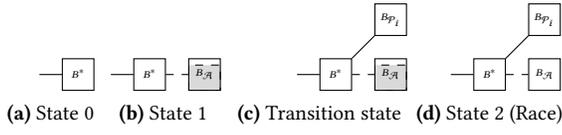
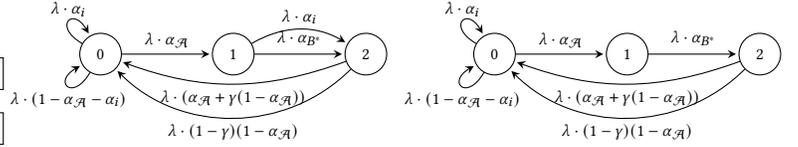

(a) State 0  (b) State 1  (c) Transition state  (d) State 2 (Race)     (a) $S_{\text{mine}}$: $\mathcal{P}_i$ mines on $B^*$ in state 1     (b) $S_{\text{stop}}$: $\mathcal{P}_i$ stops mining in state 1

**Figure 1:** *States*   **Figure 2:** *Markov chains*

On the other hand when $\mathcal{P}_i$ chooses strategy $S_{\text{mine}}$ and therefore keeps mining all the time, her cost $\hat{C}_i^{S_{\text{mine}}}$ is constant:

$$\hat{C}_i^{S_{\text{mine}}} = \lim_{t \to \infty} \frac{C_i^{S_{\text{mine}}}(t)}{t} = \alpha_i \cdot c_i.$$

Therefore, to find the more beneficial strategy, it is left to find the average revenues $\hat{R}_i^{S_{\text{stop}}}$ and $\hat{R}_i^{S_{\text{mine}}}$ for $S_{\text{stop}}$ and $S_{\text{mine}}$ respectively.

We now analyze the Markov chain: For both strategies the rational miner $\mathcal{P}_i$ receives profit $K$ every time she passes from state 0 back to state 0 with the rate $\alpha_i \lambda$ and from state 2 to 0 with rate $\alpha_i \lambda$. For strategy $S_{\text{mine}}$, $\mathcal{P}_i$ receives profit $(1-\gamma)(1-\alpha_{\mathcal{A}}) \cdot K$ when she passes from state 1 to state 2 with rate $\alpha_i \lambda$. Therefore the expected utility for strategy $S_{\text{stop}}$ is:

$$U_i^{S_{\text{stop}}} = \frac{1}{\alpha_i}(\hat{R}_i^{S_{\text{stop}}} - \hat{C}_i^{S_{\text{stop}}}) \qquad (5)$$
$$= \frac{1}{\alpha_i} \cdot ((p_0^{S_{\text{stop}}} + p_2^{S_{\text{stop}}}) \cdot \alpha_i \lambda K - (1 - p_1^{S_{\text{stop}}}) \cdot \alpha_i c_i)$$
$$= (p_0^{S_{\text{stop}}} + p_2^{S_{\text{stop}}}) \cdot \lambda K - (1 - p_1^{S_{\text{stop}}}) \cdot c_i.$$

Similarly the expected utility for strategy $S_{\text{mine}}$ is:

$$U_i^{S_{\text{mine}}} = \frac{1}{\alpha_i}(\hat{R}_i^{S_{\text{mine}}} - \hat{C}_i^{S_{\text{mine}}}) \qquad (6)$$
$$= (p_0^{S_{\text{mine}}} + p_2^{S_{\text{mine}}} + (1-\gamma)(1-\alpha_{\mathcal{A}}) \cdot p_1^{S_{\text{mine}}})\lambda K - c_i.$$

**Conditions for Successful Attack** We intend to calculate for what values of $\omega_i^b$ (defined in eq. (2)) the attack would be successful given $\alpha_{\mathcal{A}}$ and $\alpha_i$, i.e., the mining power of the attacker and a certain rational miner $\mathcal{P}_i$. Note that in order for this attack to enforce complete shutdown, we have to examine the miner with the largest mining power. Using eq. (5) and eq. (6) we define $D(\alpha_{B^*})$ to be the normalized difference between $U_i^{S_{\text{stop}}}$ and $U_i^{S_{\text{mine}}}$:

$$D(\alpha_{B^*}) \triangleq \frac{U_i^{S_{\text{stop}}} - U_i^{S_{\text{mine}}}}{c_i} \qquad (7)$$
$$= (p_0^{S_{\text{stop}}} + p_2^{S_{\text{stop}}} - p_0^{S_{\text{mine}}} - p_2^{S_{\text{mine}}}$$
$$- (1-\gamma)(1-\alpha_{\mathcal{A}}) \cdot p_1^{S_{\text{mine}}}) \cdot \omega_i^b + p_1^{S_{\text{stop}}}.$$

Our goal is to find when the attack is successful and all miners stop, that is, what are the $\omega_i^b$ values for which for all possible $\alpha_{B^*}$ values it holds that $D(\alpha_{B^*}) > 0$. We therefore calculate the condition on $\omega_i^b$ so that $D(\alpha_{B^*}) > 0$ using eq. (7):

$$\omega_i^b < \underbrace{\frac{p_1^{S_{\text{stop}}}}{p_0^{S_{\text{mine}}} + p_2^{S_{\text{mine}}} + (1-\gamma)(1-\alpha_{\mathcal{A}}) \cdot p_1^{S_{\text{mine}}} - (p_0^{S_{\text{stop}}} + p_2^{S_{\text{stop}}})}}_{Q(\alpha_{B^*})}. \qquad (8)$$

We use calculus to find the tight condition, and get that $Q(\alpha_{B^*})$ is minimal when $\alpha_{B^*} = 0$, regardless of the parameters' values.

This result implies that the motivation for a miner to keep mining during the attack decreases when other miners keep mining, as the minimum is achieved when all other miners are following $S_{\text{stop}}$. By assigning $\alpha_{B^*} = 0$ to eq. (8) and using the probabilities calculated in eq. (4), the tight condition on $\omega_i^b$ is:

$$\omega_i^b < \frac{\alpha_{\mathcal{A}} + \alpha_i + \alpha_{\mathcal{A}}\alpha_i}{\alpha_i + \alpha_{\mathcal{A}}\alpha_i + (1-\gamma)\alpha_{\mathcal{A}}(1-\alpha_{\mathcal{A}})}. \qquad (9)$$

This condition ensures that $S_{\text{stop}}$ is *dominant strategy* for $\mathcal{P}_i$. In other words, $S_{\text{stop}}$ is always the best strategy for $\mathcal{P}_i$ regardless of other payers' actions. Note that the *dominant strategy* is $S_{\text{stop}}$ for *all* miners if the condition in eq. (9) holds for all miners in the system.

### 5.2 Threshold Values

We consider specific system parameter values and the resulting threshold on $\omega_i^b$ for a successful attack.

First we use the condition on $\omega_i^b$ that was obtained in eq. (9). Figure 3 shows the highest $\omega_i^b$ that allows the attack for different values of $\alpha_{\mathcal{A}}$, $\alpha_i$ and $\gamma$. Unlike previous attacks, even an attacker with a relatively small computational power (e.g., $\alpha_{\mathcal{A}} < 0.1$) can successfully mount an attack to stop all other miners from mining. The mining power of the rational miner $\alpha_i$ is also important to the success of the attack. For example, with $\alpha_{\mathcal{A}} = 0.2$, $\gamma = \frac{1}{2}$ and $\alpha_i = 0.1$, the threshold $\omega_i^b$ is almost 1.6. Note that even if all the rational miners have similar profitability, a small attacker would be able to stop only smaller miners. This shows that larger miners are harder to attack with BDoS.

Moreover, fig. 3 shows that when $\gamma = 0$ and $\alpha_{\mathcal{A}} = 0.2$, the attacker needs $\omega_i^b$ to be smaller than 1.15 in order to attack a rational miner with $\alpha_i = 0.1$, compared to $\omega_i^b < 1.6$ when $\gamma = \frac{1}{2}$ and $\omega_i^b < 2.7$ when $\gamma = 1$. This highlights the importance of the rushing ability for the attacker. Note that $\gamma = \frac{1}{2}$ is a conservative assumption primarily since an adversary can control a relay network [50] and therefore potentially achieve $\gamma$ even closer to 1. In §8, we further show that even if the rational miners are deviating from Nakamoto's protocol by boycotting $\mathcal{A}$'s blocks (and therefore decreasing $\gamma$), she can use smart contracts (on external cryptocurrency) to make her blocks indistinguishable from rational miners' blocks.

**Fixing $\alpha_{B^*}$** We found the borderline $\omega_i^b$ for the worst case, i.e., for all possible chosen strategies of other miners. But we saw that if the portion of rational miners that keep mining $\alpha_{B^*}$ increases, the motivation for $\mathcal{P}_i$ to stop mining also increases. We now consider a scenario where $\mathcal{P}_i$ can accurately estimate $\alpha_{B^*}$. In practice, this can be done by spying on other pools [45, 112] or by monitoring the recent inter-block time. As before, we assume that $\gamma = \frac{1}{2}$. Using eq. (8), we conclude that the bound on $\omega_i^b$ is $Q(\alpha_{B^*})$. We define: $\sigma_i = \frac{\alpha_{B^*}}{1-\alpha_{\mathcal{A}}-\alpha_i}$, which is the absolute portion of rational miners other than $\mathcal{P}_i$ that continue mining. We plot the borderline $\omega_i^b$, $\alpha_{\mathcal{A}}$ and $\alpha_i$ for different $\sigma_i$ values in fig. 4.

We can see that if all other rational miners chose $S_{\text{mine}}$ ($\sigma_i = 1$), then for $\alpha_{\mathcal{A}} = 0.2$ and $\alpha_i = 0.16$, $\mathcal{P}_i$ stops mining for $\omega_i^b < 2$ which is significantly higher than $\omega_i^b < 1.45$ for the case with $\sigma_i = 0$. As expected, the threshold for a partial shutdown is significantly higher than the threshold for a complete shutdown.

## 5.3 Partial Shutdown

We now analyze partial-shutdown attacks. The total reduction in throughput the attacker can cause and the cost of the attack depend on the portion of the miners that choose $S_{\text{mine}}$.

We denote by $\sigma$ the portion of all non-adversarial miners that keep mining. The game can be described by the state machine for $S_{\text{stop}}$ we used before (fig. 2) assuming that $\alpha_i = 0$. Recall that the portion of time spent in state 1 of the $S_{\text{stop}}$ state machine (fig. 2) is $p_1$. The *throughput* is the average number of blocks that end up in the main chain per second, i.e., excluding the forked blocks. The *relative throughput* is the ratio between the throughput in the presence of the attack and the throughput in an honest game. It is easy to see that the relative throughput is $(1 - p_1)$, as the block produced in state 1 is the only block that does not contribute to the throughput. The attacker's cost is $\alpha_{\mathcal{A}} c_{\mathcal{A}} (1 - p_1)$, as the attacker mines at a constant rate in all states but 1. We define the *relative cost* as the ratio between the cost of the attacker in BDoS and her cost in an honest game. Therefore, the relative cost is $(1 - p_1)$ – the same as the relative throughput.

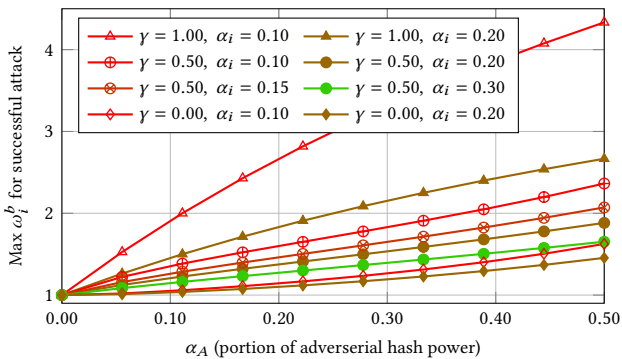

**Figure 3:** $\omega_i^b$ that will allow an attack for different $\alpha_A$, $\gamma$ and $\alpha_i$ (Notice that $\gamma$ can't reach 1 in real setting).

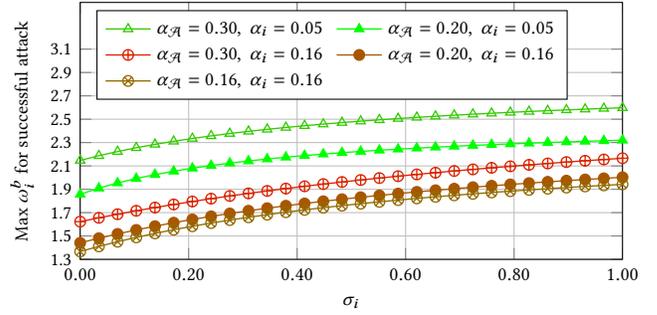

**Figure 4:** $\omega_i^b$ that will allow an attack for different $\sigma_i$, $\alpha_{\mathcal{A}}$ and $\alpha_i$ while $\gamma = \frac{1}{2}$.

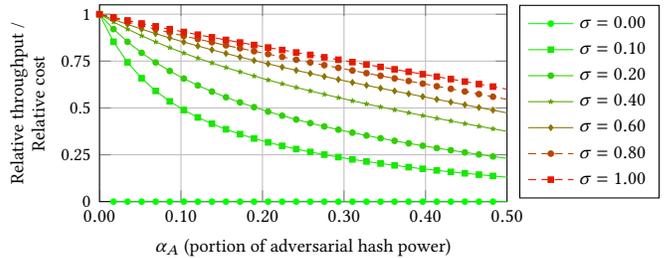

**Figure 5:** The relative throughput (out of the original coin's throughput) or relative cost for the attacker $\alpha_A$, $\sigma$.

We plot $(1 - p_1)$ vs. $\alpha_{\mathcal{A}}$ in fig. 5. With $\sigma = 0$, the growth of the blockchain completely stops, and the Markov chain remains in state 1 forever at cost 0 for the attacker. The graph indeed shows that when $\sigma$ is closer to 0, relative cost becomes smaller even for small attackers, until it collapses to 0 when $\sigma = 0$.

The result of this partial denial of service can be a panic and price crash, as cryptocurrencies' prices are subject to manipulation [2, 48]. A significant slowdown is an unusual event that can potentially start a chain reaction in the form of a price fall that can lower the profitability even further. Eventually, even a partial attack could cause irreparable damage to a cryptocurrency.

## 6 BDOS IN THE PRESENCE SPV MINING

So far, we have assumed that no rational miner mines on the block header. Note that publishing a header allows miners to extend it optimistically, as a block can be extended using only its hash. This is called *SPV mining* [97], and happens in practice. According to Nakamoto consensus, no miner considers a block that references a header as part of her main chain, as the system state is undefined without the block content. It is also impossible to validate the transactions of the next block (even if it is a full block). Therefore, when engaging in SPV mining, a miner assumes that the full block corresponding to the header will be published in the future.

In order to include the action that corresponds to SPV mining, denoted by mineSPV, we introduce few changes to our attacker algorithm, model and analysis.

## 6.1 Model Updates

To assume miners can SPV mine, we have to extend our definition for miners' behavior and assume that they can digress from the protocol with small deviations. Like Carlsten et al. [23], we say that the miners are *petty-compliant*, that is, they only take steps that almost follow the protocol, namely, (1) extending the longest chain, (2) leave the protocol or (3) engage in SPV mining. Consequently, we add a third strategy to the rational miners' strategy space we have previously considered, namely mining on the attacker's header during the attack. We denote it by $S_{\text{SPV}}$.

In addition, we assume that the $\mathcal{A}$ is aware of when a rational miner $\mathcal{P}_i$ finds a block that extends $\mathcal{A}$'s header. More formally, if $\mathcal{P}_i$ won the round, the scheduler adds her block to all other miners' ledgers, including $\mathcal{A}$'s. On a practical note, this can be done by spying on other mining pools. Thus, $\mathcal{A}$ can join all major mining pools as a miner and be warned when the mining pool manages to find a block that extends her block header.

Therefore we change $\mathcal{A}$'s strategy so that when a miner successfully finds block $B_{\mathcal{P}_i}$ that extends $B_{\mathcal{A}}$, $\mathcal{A}$ abandons $B_{\mathcal{A}}$ and return to mining on $B^*$. $\mathcal{A}$ will never publish the data that corresponds to $B_{\mathcal{A}}$, effectively invalidating $B_{\mathcal{P}_i}$.

**Note** If we consider a setting where $\mathcal{A}$ can ignore or be unaware of a new block $B_{\mathcal{P}_i}$ mined by $\mathcal{P}_i$ that extends her header $B_{\mathcal{A}}$, the attack, in fact, becomes *stronger*. This is because $B_{\mathcal{P}_i}$ would be withheld by $\mathcal{P}_i$ until $\mathcal{A}$ publishes the content of $B_{\mathcal{A}}$. But $\mathcal{A}$ would only publish it in case of a race condition, with some other block $B_{\mathcal{P}_j}$. Once $\mathcal{A}$ publishes $B_{\mathcal{A}}$, $\mathcal{P}_i$ immediately publishes $B_{\mathcal{P}_i}$, causing $\mathcal{A}$ to win the race and to invalidate $B_{\mathcal{P}_j}$. So if miners choose $S_{\text{SPV}}$ in such a setting, they only decrease the motivation for other miners to choose $S_{\text{mine}}$. Moreover, the blocks mined with SPV are likely to be empty, as a miner who does not possess all the transactions in the current state would not risk invalidating her block by causing conflicts. We, therefore, regard the analysis of this case as being outside the scope of this paper.

## 6.2 BDoS Effectiveness in the Presence of $S_{\text{SPV}}$

After updating the model and adversarial strategy, it remains to show that the effectiveness of the attack does not decrease if some portion of the miners engages in SPV mining. The intuitive argument is that if a rational miner decides to SPV mine, her next block would not be included in the main chain. Therefore, the utility of $S_{\text{mine}}$ is strictly higher than the utility of $S_{\text{SPV}}$. Consequently, assuming that all miners are rational, no miner would choose to engage in SPV mining. The quantitative analysis (§5) of the conditions for the attack still holds. The formal justification of this argument is more complicated and is based on a different solution concept – Iterated Elimination of Strictly Dominated Strategies (IESDS). We defer the detailed analysis to Appendix D.

## 7 TWO-COIN MODEL

So far, we used a model where the attacker initiates an attack on a cryptocurrency, and the rational miners can either mine on this coin or not mine at all.

We now consider a *two-coin model* where miners can choose to mine on one of two coins. This requires the two coins to employ a similar mining algorithm so that miners could mine on both coins with similar efficiency. The main difference from the previous model is that miners have less to lose than by ceasing mining completely. If the profitability of the coins is similar, even if the attacker lowers the expected profit slightly, the miners would still be motivated to switch to the other coin.

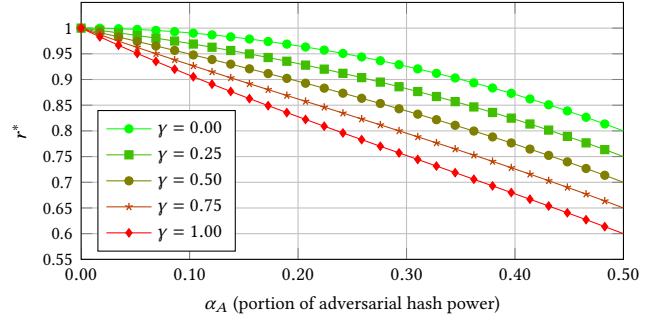

**Figure 6:** $r^*$ that will allow the attack

Due to the large number of coins in the blockchain world and the fact that some of them use the same or similar mining schemes [53, 58, 69, 81, 96, 104, 109, 113], the more vulnerable two-coin model is in fact more realistic as well.

When there is a profitability difference, miners are expected to switch coins to the more profitable coin. By doing that, they cause the profitability to decrease in the long term (due to difficulty adjustment) and bring the coins' profitability to equilibrium. BDoS results in artificial profitability difference between the coins, causing all rational miners to abandon the attacked coin for the other coin.

### 7.1 Model Changes

In our two-coin model, we assume a rational miner can choose between mining on $C_1$ or a competing coin $C_2$ with the same mining mechanism. We denote the profitability and utility of $\mathcal{P}_i$ for coin $C_1$ with $\omega_i^b$ and $U_i$ respectively, and the profitability and utility of $\mathcal{P}_i$ for coin $C_2$ with $\omega_i^{b'}$ and $U_i'$ respectively. If the honest setting profitability for the miner $\mathcal{P}_i$ on both coins is equal $\omega_i^b = \omega_i^{b'}$, then the attacker no longer has a threshold for an attack on coin $C_1$. This is because the attack always decreases the mining utility $U_i$ for $\mathcal{P}_i$ who mines on coin $C_1$ and therefore every miner would choose to mine on $C_2$ instead (as $U_i^{b'} > U_i$).

### 7.2 Analysis

The model is almost the same as the one described in section 3.2, and the analysis would be similar to the analysis in §5. The main difference is that we no longer consider a choice between mine on $B^*$ and stop but between mining on $B^*$ in the attacked coin $C_1$ ($S_{mine}^1$) and mining on another coin $C_2$ ($S_{mine}^2$). The utility $U_i^{S_{mine}^1}$ for $\mathcal{P}_i$ for the first strategy $S_{mine}^1$ is the same as $U_i^{S_{\text{mine}}}$ in eq. (6), thus $U_i^{S_{mine}^1} = U_i^{S_{\text{mine}}}$.

While the utility $U_i^{S_{mine}^2}$ for $\mathcal{P}_i$ for the second strategy $S_{mine}^2$ (mining in the honest setting in coin $C_2$) is similar to $U_i^b$ in eq. (1). We use different $\lambda$, $c_i$ and $K$ parameters for the second coin ($\lambda'$, $c_i'$ and $K'$ respectively), as they are not necessarily the same for both coins. Thus: $U_i^{S_{mine}^2} = U_i^{b'} = \lambda' K' - c_i'$. To compare the two utilities in two different coins, we can no longer use the normalized utility, as the mining power constants $\alpha_i$ and $\alpha_i'$ of coin $C_1$ and $C_2$ respectively, are not necessarily the same. Note that the mining cost per second of $\mathcal{P}_i$ is equal for both coins, so that $\alpha_i c_i = \alpha_i' c_i'$. We define $D$ as the difference between the two utilities $U_i^{S_{mine}^1}$ and $U_i^{S_{mine}^2}$, when each utility is multiplied by the respective hashrate:

$$D(\alpha_{B^*}) \triangleq \frac{\alpha_i U_i^{S_{mine}^1} - \alpha_i' U_i^{S_{mine}^2}}{\alpha_i c_i} = \frac{\alpha_i U_i^{S_{mine}^1}}{\alpha_i c_i} - \frac{\alpha_i' U_i^{S_{mine}^2}}{\alpha_i' c_i'}$$
$$= (p_0^{S_{mine}} + p_2^{S_{mine}} + (1-\gamma)(1-\alpha_{\mathcal{A}}) \cdot p_1^{S_{mine}}) \cdot \omega_i^b - \omega_i^{b'}.$$

We ask when it holds that $D(\alpha_{B^*}) > 0$, thus look for the ratio $r$ s.t:

$$r \triangleq \frac{\omega_i^{b'}}{\omega_i^b} > \underbrace{(p_0^{S_{mine}} + p_2^{S_{mine}} + (1-\gamma)(1-\alpha_{\mathcal{A}}) \cdot p_1^{S_{mine}})}_{W(\alpha_{B^*})}.$$

We now need to calculate the maximal value $W(\alpha_{B^*})$ can get. Using calculus we derive that it attains maximum for $\alpha_{B^*} = 1 - \alpha_{\mathcal{A}} - \alpha_i$ which holds when all other miners do not switch coins (as this is the maximum utility they can get):

$$r > W(\alpha_{B^*} = 1 - \alpha_{\mathcal{A}} - \alpha_i) = \frac{(1-\alpha_{\mathcal{A}})(\alpha_{\mathcal{A}}(\gamma-2)-1)}{\alpha_{\mathcal{A}}^2 - \alpha_{\mathcal{A}} - 1} = r^*.$$

An interesting fact is that the minimal $r$ that allows the attack, which we denote by $r^*$, does not depend on $\mathcal{P}_i$'s mining power.

We plot $r^*$ that allows the attack for different $\gamma$ and $\alpha_{\mathcal{A}}$ in fig. 6. When $\gamma = \frac{1}{2}$ and $\alpha_{\mathcal{A}} = 0.2$, it holds that $r^* = 0.9$. This means such an attacker can launch the attack as long as the profitability of $C_1$ is smaller by at least 11% than the profitability of $C_2$. Note that the attack is always possible when the profitability of $C_2$ is equal to the one of $C_1$, i.e. $r = 1$.

### 7.3 Estimating Practical $r$

It is shown in [69] that such migrations between coins happen frequently. In [113], the authors found a correlation between profitability changes of Bitcoin and Bitcoin Cash with the hashrate changes of the two coins, concluding that miners migrate between coins according to profitability.

The ratio between profitabilities of Bitcoin and Bitcoin Cash, presented in [35], is equivalent to $r$ that was defined in §7. We can see that $r$ is close to 1 and often is within 5% error range. Such realistic values for $r$ would allow an attacker with 10% of the mining power to mount a successful attack.

## 8 PRACTICAL CONCERNS

Until now we analyzed BDoS within the model introduced in §3. In this section, we discuss further the practical aspects of launching BDoS. For this purpose, we use Bitcoin as a case study coin.

### 8.1 Motivation of the Attacker

The attacker described in this work is different from profit-driven attackers (or deviating miners) [45, 46, 68, 89, 102]. In the case of BDoS, the attacker has exogenous motivation for the attack. We now give some examples of such attackers:

**Goverment** The rise of cryptocurrencies has raised concern in some states regarding the negative effects on the strength of their own fiat money and monetary policy effectiveness [9, 11, 54, 80]. Cyberspace allows large actors to achieve a political or economic goal while maintaining plausible deniability. Previous reports suggested that governments deploy costly and elaborate attacks to achieve such goals while hiding their traces [27, 70, 78, 105].

Note that although BDoS requires the attacker to announce the attack publicly, it does not require the attacker to reveal her real identity. The Nakamoto consensus allows a mining farm to have minimal exposure to the network and it only needs to receive previous blocks' hashes and send the data of the new block when it is found. Therefore, the attacker's P2P node can be a server with no link to discernible its operator. Governments reportedly [30, 78, 94] use frontman or proxy entities regularly.

Therefore, given the high motivation to destabilize cryptocurrencies, the ability to allocate significant resources for this goal, and the possibility of carrying out the attack covertly, state actors could attempt to execute BDoS in the future.

**External profit** An attacker can gain profit by taking a short position on a cryptocurrency, and then sabotaging the coin to cause a price collapse [20, 101]. Such an attacker can potentially gain even more by conducting a partial attack, as her aim is not to stop the coin completely but temporarily destabilize it.

**Competing coins** The attacker could be affiliated with a competing coin, and willing to spend resources to weaken her competitor. There is evidence that 51% attacks (much more expensive than BDoS) have aimed to sabotage competing coins [75, 93].

### 8.2 Rationality of Miners

So far we assumed that all the miners are rational. We would like to understand what is the effect of non-rational (or altruistic) miners on the success of the attack.

First, on a practical level, there is evidence that miners tend to stop mining when it is no longer profitable [28, 43]. Moreover, miners tend to switch among competing coins due to differing profitability [69, 113]. This switching happens when mining is still profitable on both coins, and strongly suggests those miners are acting greedily, and not altruistically supporting a particular coin.

Second, our model can describe a system with altruistic miners. We can consider them by assuming that some miners have $0 electricity cost. In the presence of altruistic miners, the attacker would not cause a *complete shutdown* but rather a *partial shutdown*. A partial shutdown attack can be the desired result for certain attackers (section 8.1), because it can destabilize the coin and cause the price to crash, yet still allow transactions to be approved.

Finally, the security of blockchains should not rely on altruism [108]. The original Nakamoto consensus [88] was designed to compensate miners for their work and consequently incentivize them to stay honest. Therefore, blockchain security analysis should consider the worst-case when *all miners* decide to mine based on the mining profitability [43, 108]. Otherwise, security is contingent on the existence of miners that are willing to pay to keep the coin alive. This is a strong requirement for financial systems with a total market cap of hundreds of billions of dollars.

### 8.3 Header-Publication Method

BDoS requires the adversary to announce a block header without revealing a full block. We now describe two publishing methods that the adversary can use in practice. Crucially, regardless of the chosen header publication method, a rational miner would never bury her head in the sand, i.e., ignore the additional info and hence reduce her profit. In practice, reports [28, 43, 69] suggest that miners indeed poll external data and change their strategy according to it, e.g., mine on a more profitable coin or stop mining completely.

In the first method, the attacker announces that she is committing to an attack, and publishes her block headers on a dedicated web page. Rational miners would poll this web page.

In the second publication method, the attacker directly sends block headers (of the block she created) to other miners. In Bitcoin, this can be done using the built-in BIP152 protocol [31]. However, the protocol dictates that nodes should not propagate a block header until fully validating the entire block. Thus, the attacker must be connected directly to other miners. As before, to maximize her utility, a rational miner would not ignore such headers.

### 8.4 Alternative Proofs of Block Possession

Instead of sending a block header, the adversary can use a smart contract (potentially on an external blockchain) to demonstrate the discovery of a block without revealing its header.

The idea is to use an *economic mechanism* to demonstrate knowledge of a valid block header $H$. Briefly, the attacker places large collateral in the contract, possibly on a different cryptocurrency like Ethereum, along with a cryptographic commitment *Comm* to $H$ and with the hash of the previous block. If at some predetermined (distant) future time, she de-commits a valid $H$ for the contract, i.e., one that points to the previous block, she recovers the collateral. Otherwise, she forfeits the collateral to miners. If the contract is part of a different blockchain, miners can be required to prove their identity using a signature with their private key from the attacked coin. Thus, the attacker is incentivized to claim and commit only to a valid header, but *need not reveal any information about $H$*, until $H$ is no longer useful to miners.

To ensure that the attacker has the incentive to commit to a valid $H$, the collateral should be significantly larger than the cost of mining blocks during the commitment period. The collateral, if forfeited, can be split among a predefined list of mining pools (weighted by their mining power). For example, this list might include miners of the last, e.g., 1,000 blocks.

This approach has one key advantage over the block header approach: until $H$ is de-committed (again, in the far future), no rational miner can distinguish the attacker's block from an honest block as, during the race, the other miners only posses a commitment for the block rather than a block header or hash. This approach prevents rational miners from forming a coalition that would ignore the attacker's block. When the attacker reveals the block after a long time, it would be impractical to ignore it, as it would have a large number of confirmations [3].

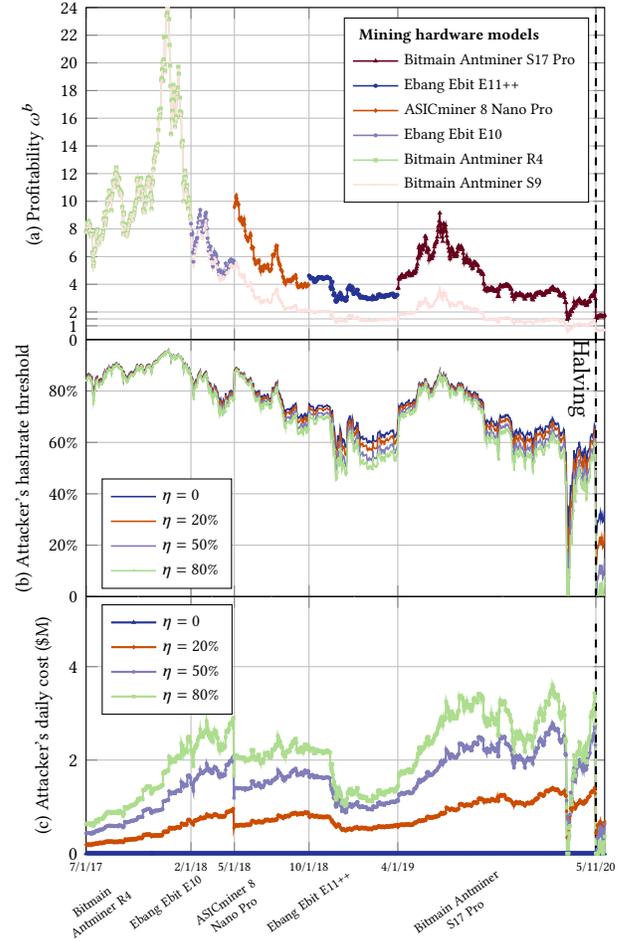

**Figure 7:** *(a) Profitability of mining Bitcoin; (b) Attacker's resource threshold; (c) Attacker's daily cost. The hops indicated by the vertical grids result from mining hardware switches due to release of new profitable models.*

Another method that can achieve a similar effect is Zero-Knowledge proof. An attacker can publish a non-interactive Zero-Knowledge proof on her website and prove she found a block header without exposing identifying information like the block hash. Like in the case of smart contracts, rational miners cannot distinguish the attacker's block from an honest block in case of a race.

Exact details for both methods are beyond the scope of this paper.

### 8.5 Practical $\omega^b$

The success of the attack relies critically on the baseline profitability $\omega^b$. To estimate realistic values for $\omega^b$, we study the properties of Bitcoin, as the archetypal PoW cryptocurrency. First, we consider the costs that affect $\omega^b$. Next, we explore how and when $\omega^b$ changes in practice. This is important due to the attacker's liberty to choose the moment of the attack. Finally, we estimate real values for $\omega^b$, using both previous work and our own analysis.

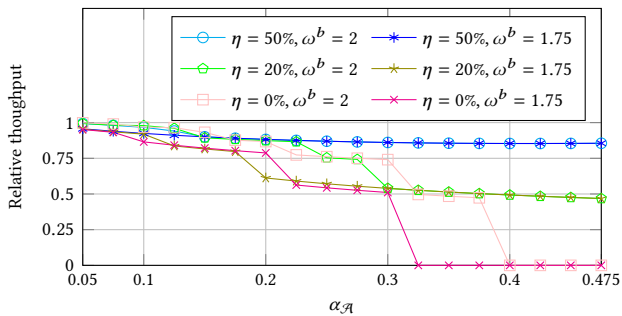

**Figure 8:** *Relative throughput in the Nash equilibrium.*

**CAPEX and OPEX** The miner's expenses include operating expenses (OPEX or ongoing cost) and capital expenses (CAPEX or sunk cost) [53, 64]. OPEX includes electricity and cooling. CAPEX includes costs like buying/renting facilities and mining hardware. As we compare $S_{\text{stop}}$ to other strategies, we can ignore the CAPEX element because all strategies have an identical initial cost. Moreover, we showed (§5) that the CAPEX is not relevant in our infinite-horizon game, as it does not change the profit per second.

Nevertheless, high CAPEX affects $\omega^b$ indirectly, as miners have to return their initial investment. Lower CAPEX can cause $\omega^b$ to decrease as more potential miners would join the game, bringing the system closer to equilibrium [13, 52, 59, 117], i.e., $\omega^b$ close to 1. As we showed in §5 this would hurt the security of the system as small attackers would be able to mount a successful BDoS attack.

**$\omega^b$ Fluctuations** More than 50% of Bitcoin's hash power is located in Sichuan, China [126], since the electricity price there is extremely low during the wet season – as low as $0.04 per kWh [13], slightly varying between hydropower plants. Although it is easier for a BDoS attacker to target miners in regions with higher electricity prices (fig. 3), we show below (section 8.6) that miners in Sichuan (with minimal electricity prices) are still at risk. Moreover, at the end of the wet season or in an unexpected dry period [55], before the difficulty adjustment, the electricity prices increase, therefore the profitability of most miners can be exceptionally low. This is an ideal moment for an attacker to attack as $\omega^b$ is expected to be at its lowest point.

Another essential factor that can make the system vulnerable to BDoS is the block reward adjustment that happened in May 2020 [10]. The block reward has dropped from 12.5BTC to 6.25BTC. The transition was immediate and halved $\omega^b$.

**Estimating Upper Bound for $\omega^b$** Estimating $\omega^b$ is based on several parameters. Mining hardware rates and electricity consumption of different mining hardware are available in *ASIC Miner Value* [84]. We analyzed Bitcoin blocks 471744 to 624960 (June 17, 2017 – April 8, 2020) using the Google BigQuery [37] Bitcoin dataset to collect mining difficulty data and compute the expected number of hashes needed to find a block [119]. We consider an electricity price of $0.04 per kWh with additional 15% cooling & other OPEX cost [13] and Bitcoin prices from [32]. We show in fig. 7(a) the profitability of mining Bitcoin with the best hardware at each point in time as well as with Antminer S9 which was prevalent until 2019.

The dashed vertical line indicates the most recent Bitcoin block reward halving [10]. We can see that $\omega^b$ has halved immediately.

The lowest historical threshold on adversarial hashrate for a complete shutdown BDoS, derived according to eq. (9), when the largest rational miner has 20% of hashing power and $\gamma = \frac{1}{2}$ was 21% on March 13, 2020. After the Bitcoin block reward halving that took place in May 11, 2020 [10], the average threshold for the attack has reduced significantly and the complete shutdown threshold ranged between 24% and 34% until Jun 5, 2020 (fig. 7(b)). Note that the attacker can choose the timing of the attack and therefore we are interested in finding the minimal threshold rather than the average threshold for the attack.

This attack threshold is significantly lower than the previously known bound for a complete shutdown attack which is 51%.

Our estimation for electricity cost is conservative and other sources estimate it to be higher on average [12, 39, 43, 92, 127]. For less conservative electricity prices, the attack is significantly cheaper. The threshold for the attacker's hash rate was as low as 7.7% before halving and 10% after halving for $0.05/kWh. It was less than 1% for both before and after halving for $0.06/kWh. We give more details about the attack threshold for other electricity prices in (Appendix F).

## 8.6 Partial Attack and Its Costs

A BDoS attacker can stop rational miners from mining and cause a significant slowdown to the system, even when altruistic miners exist (section 5.3). We use Bitcoin as a case study and estimate the mining power threshold and daily cost for the attacker.

We calculate the attack resource threshold over time to stop all rational miners in Bitcoin based on our estimated mining profitability data (fig. 7(a)). We assume $\mathcal{A}$'s rushing factor $\gamma$ is $\frac{1}{2}$. We denote by $\eta$ the proportion of mining power owned by altruistic miners among non-adversarial miners. Recall that the mining power required to stop the largest rational miner suffices to stop all rational miners (fig. 3). To be conservative, we assume the largest rational miner controls 20% of the mining power [19]. As before, we consider the adversary as a new miner entering the network in addition to existing miners, so the actual mining power of each rational miner in the attack is scaled by a factor $(1 - \alpha_{\mathcal{A}})$. We assume both the attacker and the rational miners use the most profitable mining hardware available at each moment; thus, they have the same profitability. It is easier to attack when rational miners know the value of $\eta$ than when they don't because the attack threshold decreases with the ratio of miners remaining mining (fig. 4). We assume each rational miner knows $\eta$ for a tighter bound on attack threshold. An attacker can achieve this by first attacking with a lot of hash power for a short while to stop all rational miners and estimate $\eta$ according to the reduction in the mining rate.

We calculate and plot the resource threshold required for a BDoS attacker (fig. 7(b)). As of March 13, 2020, 21% mining power suffices to launch BDoS when there are no altruistic miners, and even less than 10% when there are at least 20% altruistic miners.

In addition, we estimate the attacker's daily OPEX cost (fig. 7(c)). As of March 13, 2020, an attacker would spend less than $1M daily to launch BDoS. When all miners are rational, the attacker's OPEX cost is zero. For comparison, the cost per day of a 51% attack would

have been $10.6M with Antminer S9 SE [90] rigs and $5.2M with Antminer S17 Pro [15] (see Appendix C for details).

## 8.7 Simulation of Realistic Miners

Even if the mining profitability is not low enough to stop all rational miners, a BDoS attacker can still discourage some small miners from mining. This is because the upper bound on $\omega_i^b$, that drives a rational miner $\mathcal{P}_i$ out of mining, increases with the mining power of $\mathcal{P}_i$ (fig. 3). Assuming all miners have the same profitability factor $\omega^b$, consider an adapting process in which rational miners are always aware of the total mining power of active miners in the network, and choose between $S_{\text{mine}}$ and $S_{\text{stop}}$ adaptively. A rational miner $\mathcal{P}_i$ chooses to mine if the real-world $\omega^b$ is higher than the upper bound for $\mathcal{P}_i$, and stops mining otherwise. The process starts with the state in which everyone mines and ends when no rational miner changes her strategy anymore. Thus at the end of the process, rational miners reach a Nash equilibrium.

We simulate the process for Bitcoin to find an equilibrium for all rational miners. As before, we assume $\gamma = \frac{1}{2}$ and use the mining power distribution as of March 2020 [19] for rational miners. We assume $\eta$ of the non-adversarial mining power belongs to altruistic miners, thus the hash power of each rational miner is scaled by a factor $(1 - \eta)$.

We plot the resulting relative throughput in the Nash Equilibrium with different mining profitability $\omega^b$ and altruistic parameter $\eta$ in separate curves in fig. 8. When $\alpha_{\mathcal{A}}$ is low, only some small miners choose $S_{\text{stop}}$. As $\alpha_{\mathcal{A}}$ grows, it becomes profitable for more medium-sized miners to stop mining. The simulation indicates that there is a wide range of profitability factor values that allow BDoS that causes a significant slowdown in practice. For example, with only 20% of mining power, an attacker may slow down the system by 40% even when the real-life profitability is 1.75 and 20% of the non-adversarial miners are altruistic.

## 9 MITIGATION

We now describe possible mitigations for BDoS attacks.

**Uncle blocks** BDoS is designed against Nakamoto consensus cryptocurrencies like Bitcoin. However, it is not effective against other cryptocurrencies, e.g., Ethereum [22, 123]. This is due to the uncle *block mechanism* [123], which rewards miners who mined blocks that are directly connected to the main chain. This thwarts BDoS, as now, in case a rational miner loses the race, her block is still rewarded. Thus, by publishing a block header, the attacker no longer reduces the expected profit of rational miners as significantly.

Note that the mechanism does not grant a reward for blocks that are not directly connected to the main chain. Consequently, there may be similar attacks that still allow the attacker to decrease the expected reward, e.g., by publishing two-block headers that fork the most recent block in the chain. However, the design for such an attack is beyond the scope of this work.

**Ignoring attacker's block during race** Another possible way to weaken the attack is to change miner behavior so that if there is a fork, a miner should prefer blocks not generated by an attacker. The challenge is to identify attack blocks. A third party service for this goal is out of the question as it violates the decentralized nature of the system and allows false incrimination. Instead, we propose to classify according to the time interval between the reception of the header and the reception of the block. We can safely assume that for a non-attack block, this interval is bounded by, e.g., one minute, and blocks with a longer interval are suspect.

Note that this mitigation is possible only when the adversary chooses to prove that she mined a block using a block header. This solution does not work with other methods like smart contracts and ZK proofs (section 8.4).

## 10 CONCLUSION

We present BDoS, the first Blockchain denial-of-service attack that uses incentive manipulation. BDoS sabotages the incentive mechanism behind Nakamoto's consensus by proving the attacker has achieved an advantage in mining without releasing her complete block. Such proof reduces miners' incentive to mine, making it less profitable than not mining. Thus, rational profit-driven miners would cease mining. We show that cryptocurrencies based on Nakamoto's blockchain are vulnerable to BDoS under realistic settings, and propose mitigations.

The header-only publication capability we present is a realistic extension of the standard model under which blockchain protocols are typically analyzed. This opens the door to study new equilibria and strategies where a miner manipulates the system to increase her revenue rather than sabotage the system.

Additionally, BDoS applies to heaviest-chain PoW blockchains such as Bitcoin, Litecoin, Bitcoin-Cash, Zcash, and others. We defer to future work the questions of whether there are similar attacks against other protocols like Ethereum and whether our heuristic mitigation applies there as well.

Finally, alternative incentive-based DoS attacks may exist, possibly more efficient than BDoS. The security of blockchain protocols relies on finding the general bounds for such attacks, as well as mitigations.

## 11 ACKNOWLEDGMENT

We thank the anonymous reviewers and our shepherd, Yongdae Kim, for their feedback and suggestions. This project received funding from ISF grant 1641/18; NSF grants CNS-1514163, CNS-1564102, and CNS-1704615; BSF; the Technion Cyber-Security Center; a Mel Berlin Scholarship; a Bloomberg Fellowship; Engima MPC; as well as support from IC3 industry partners.

# REFERENCES

[1] 2016. [release/1.3.4] core: Added new TD strategy which mitigate the risk f… · ethereum/go-ethereum@bcf5657. https://github.com/ethereum/go-ethereum/commit/bcf565730b1816304947021080981245d084a930

[2] 2018. Bitcoin Price Crash Caused by Panic Sellers and Manipulation (Not Coinrail). https://bitcoinist.com/bitcoin-price-panic-sellers-manipulation-coinrail/

[3] 2020. Confirmation - Bitcoin Wiki. https://en.bitcoin.it/wiki/Confirmation

[4] Maria Apostolaki, Gian Marti, Jan Müller, and Laurent Vanbever. 2018. SABRE: Protecting Bitcoin against Routing Attacks. *arXiv preprint arXiv:1808.06254* (2018).

[5] Maria Apostolaki, Aviv Zohar, and Laurent Vanbever. 2017. Hijacking bitcoin: Routing attacks on cryptocurrencies. In *2017 IEEE Symposium on Security and Privacy (SP)*. IEEE, 375–392.

[6] Robert J Aumann. 1976. Agreeing to disagree. *The annals of statistics* (1976), 1236–1239.

[7] Christian Badertscher, Peter Gaži, Aggelos Kiayias, Alexander Russell, and Vassilis Zikas. 2018. Ouroboros genesis: Composable proof-of-stake blockchains with dynamic availability. In *Proceedings of the 2018 ACM SIGSAC Conference on Computer and Communications Security*. ACM, 913–930.

[8] Qianlan Bai, Xinyan Zhou, Xing Wang, Yuedong Xu, Xin Wang, and Qingsheng Kong. 2018. A Deep Dive into Blockchain Selfish Mining. *arXiv preprint arXiv:1811.08263* (2018).

[9] Billy Bambrough. 2019. Bitcoin Threatens To 'Take Power' From The U.S. Federal Reserve. https://www.forbes.com/sites/billybambrough/2019/05/15/a-u-s-congressman-is-so-scared-of-bitcoin-and-crypto-he-wants-it-banned/

[10] Billy Bambrough. 2020. Bitcoin Has Halved—What Now? https://www.forbes.com/sites/billybambrough/2020/05/12/bitcoin-has-halved-what-now/

[11] Billy Bambrough. 2020. The U.S. Is Very Worried About Bitcoin—And It's Finally Doing Something About It. https://www.forbes.com/sites/billybambrough/2020/02/18/the-us-is-very-worried-about-bitcoinand-its-finally-doing-something-about-it/

[12] Christopher Bendiksen, Samuel Gibbons, and E Lim. 2018. The Bitcoin Mining Network-Trends, Marginal Creation Cost, Electricity Consumption & Sources. *CoinShares Research* 21 (2018).

[13] Christopher Bendiksen, Samuel Gibbons, and E Lim. 2019. The Bitcoin Mining Network-Trends, Marginal Creation Cost, Electricity Consumption & Sources. *CoinShares Research* (2019).

[14] Billfodl. 2020. bitcoinfees. https://billfodl.com/pages/bitcoinfees

[15] Bitmain. 2019. https://shop.bitmain.com/product/detail?pid=00020191023161655 4895rHmxLOT06C2

[16] Joseph Bonneau. 2018. Hostile blockchain takeovers (short paper). In *International Conference on Financial Cryptography and Data Security*. Springer, 92–100.

[17] Joseph Bonneau, Edward W Felten, Steven Goldfeder, Joshua A Kroll, and Arvind Narayanan. 2016. Why buy when you can rent? bribery attacks on bitcoin consensus. (2016).

[18] Danny Bradbury. 2013. Feathercoin hit by massive attack. https://www.coindesk.com/feathercoin-hit-by-massive-attack

[19] BTC.com. 2020. Pool Distribution. https://btc.com/stats/pool?pool_mode=month

[20] Eric Budish. 2018. *The economic limits of bitcoin and the blockchain*. Technical Report. National Bureau of Economic Research.

[21] Vitalik Buterin. 2018. Discouragement Attacks. (2018). https://github.com/ethereum/research/blob/master/papers/discouragement/discouragement.pdf

[22] Vitalik Buterin et al. 2014. A next-generation smart contract and decentralized application platform. *white paper* 3 (2014), 37.

[23] Miles Carlsten, Harry Kalodner, S Matthew Weinberg, and Arvind Narayanan. 2016. On the instability of bitcoin without the block reward. In *Proceedings of the 2016 ACM SIGSAC Conference on Computer and Communications Security*. ACM, 154–167.

[24] Miguel Castro, Peter Druschel, Ayalvadi Ganesh, Antony Rowstron, and Dan S Wallach. 2002. Secure routing for structured peer-to-peer overlay networks. *ACM SIGOPS Operating Systems Review* 36, SI (2002), 299–314.

[25] Ethan Cecchetti, Ian Miers, and Ari Juels. 2018. PIEs: Public Incompressible Encodings for Decentralized Storage. *IACR Cryptology ePrint Archive* 2018 (2018), 684.

[26] Lin Chen, Lei Xu, Nolan Shah, Zhimin Gao, Yang Lu, and Weidong Shi. 2017. On security analysis of proof-of-elapsed-time (poet). In *International Symposium on Stabilization, Safety, and Security of Distributed Systems*. Springer, 282–297.

[27] Thomas M Chen and Saeed Abu-Nimeh. 2011. Lessons from stuxnet. *Computer* 44, 4 (2011), 91–93.

[28] CoinDesk. 2019. Bitcoin Mining Power Sees Short-Term Drop as Rainy Season Ends in China. https://www.coindesk.com/bitcoin-mining-power-sees-short-term-fallback-as-rainy-season-ends-in-china

[29] CoinMarketCap. 2019. Cryptocurrency Market Capitalizations. https://coinmarketcap.com/

[30] Michael Connell and Sarah Vogler. 2017. *Russia's Approach to Cyber Warfare (1Rev)*. Technical Report. Center for Naval Analyses Arlington United States.

[31] Matt Corallo. 2016. BIP 152: compact block relay. See *https://github.com/bitcoin/bips/blob/master/bip-0152.mediawiki* (2016).

[32] CryptoDataDownload. 2020. Kraken exchange data. https://www.CryptoDataDownload.com

[33] Philip Daian, Steven Goldfeder, Tyler Kell, Yunqi Li, Xueyuan Zhao, Iddo Bentov, Lorenz Breidenbach, and Ari Juels. 2019. Flash Boys 2.0: Frontrunning, Transaction Reordering, and Consensus Instability in Decentralized Exchanges. *arXiv preprint arXiv:1904.05234* (2019).

[34] Phil Daian, Rafael Pass, and Elaine Shi. 2019. Snow White: Robustly Reconfigurable Consensus and Applications to Provably Secure Proof of Stake. In *International Conference on Financial Cryptography and Data Security*. Springer, 23–41.

[35] Coin Dance. 2020. Daily Bitcoin Cash Profitability Against Bitcoin. https://cash.coin.dance/blocks/profitability

[36] Bernardo David, Peter Gaži, Aggelos Kiayias, and Alexander Russell. 2018. Ouroboros praos: An adaptively-secure, semi-synchronous proof-of-stake blockchain. In *Annual International Conference on the Theory and Applications of Cryptographic Techniques*. Springer, 66–98.

[37] Allen Day and Colin Bookman. 2018. Bitcoin in BigQuery: blockchain analytics on public data. https://cloud.google.com/blog/products/gcp/bitcoin-in-bigquery-blockchain-analytics-on-public-data

[38] Matthew De Silva. 2019. Ethereum Classic is under attack. https://qz.com/1516094/ethereum-classic-got-hit-by-a-51-attack/

[39] Oscar Delgado-Mohatar, Marta Felis-Rota, and Carlos Fernández-Herraiz. 2019. The Bitcoin mining breakdown: Is mining still profitable? *Economics Letters* 184 (2019), 108492.

[40] Department of Homeland Security. 2018. Understanding Denial-of-Service Attacks. https://www.us-cert.gov/ncas/tips/ST04-015

[41] John R Douceur. 2002. The sybil attack. In *International workshop on peer-to-peer systems*. Springer, 251–260.

[42] Cynthia Dwork and Moni Naor. 1992. Pricing via processing or combatting junk mail. In *Annual International Cryptology Conference*. Springer, 139–147.

[43] Aryaz Eghbali and Roger Wattenhofer. 2019. 12 Angry Miners. In *Data Privacy Management, Cryptocurrencies and Blockchain Technology*. Springer, 391–398.

[44] Shayan Eskandari, Seyedehmahsa Moosavi, and Jeremy Clark. 2019. SoK: Transparent Dishonesty: front-running attacks on Blockchain. (2019).

[45] Ittay Eyal. 2015. The miner's dilemma. In *2015 IEEE Symposium on Security and Privacy*. IEEE, 89–103.

[46] Ittay Eyal and Emin Gün Sirer. 2018. Majority is not enough: Bitcoin mining is vulnerable. *Commun. ACM* 61, 7 (2018), 95–102.

[47] Drew Fudenberg and Jean Tirole. 1991. Game theory, 1991. *Cambridge, Massachusetts* 393, 12 (1991), 80.

[48] Neil Gandal, J. T. Hamrick, Tyler Moore, and Tali Oberman. 2017. Price manipulation in the Bitcoin ecosystem. https://voxeu.org/article/price-manipulation-bitcoin-ecosystem

[49] Juan Garay, Aggelos Kiayias, and Nikos Leonardos. 2015. The bitcoin backbone protocol: Analysis and applications. In *Annual International Conference on the Theory and Applications of Cryptographic Techniques*. Springer, 281–310.

[50] Arthur Gervais, Ghassan O Karame, Karl Wüst, Vasileios Glykantzis, Hubert Ritzdorf, and Srdjan Capkun. 2016. On the security and performance of proof of work blockchains. In *Proceedings of the 2016 ACM SIGSAC conference on computer and communications security*. ACM, 3–16.

[51] Yossi Gilad, Rotem Hemo, Silvio Micali, Georgios Vlachos, and Nickolai Zeldovich. 2017. Algorand: Scaling byzantine agreements for cryptocurrencies. In *Proceedings of the 26th Symposium on Operating Systems Principles*. ACM, 51–68.

[52] Guy Goren and Alexander Spiegelman. 2019. Mind the Mining. *arXiv preprint arXiv:1902.03899* (2019).

[53] Adam S Hayes. 2017. Cryptocurrency value formation: An empirical study leading to a cost of production model for valuing bitcoin. *Telematics and Informatics* 34, 7 (2017), 1308–1321.

[54] Dong He. 2018. Monetary policy in the digital age: Crypto assets may one day reduce demand for central bank money. *A Quarterly Publication of the International Monetary Fund* 55, 2 (2018), 20–21.

[55] WIll Heasman. 2019. Bitcoin's Difficulty Falls As Miners Capitulate; Will They Survive the Halving? https://www.ccn.com/bitcoin-difficulty-falls-miners-capitulate/

[56] Ethan Heilman, Alison Kendler, Aviv Zohar, and Sharon Goldberg. 2015. Eclipse attacks on bitcoin's peer-to-peer network. In *24th {USENIX} Security Symposium ({USENIX} Security 15)*. 129–144.

[57] Alyssa Hertig. 2019. Bitcoin Cash Miners Undo Attacker's Transactions With '51% Attack'. https://www.coindesk.com/bitcoin-cash-miners-undo-attackers-transactions-with-51-attack

[58] MINING POOL HUB. 2019. MINING POOL HUB. https://miningpoolhub.com/

[59] Gur Huberman, Jacob Leshno, and Ciamac C Moallemi. 2019. huberman2019economic. *Columbia Business School Research Paper* 17-92 (2019).

[60] Markus Jakobsson and Ari Juels. 1999. Proofs of work and bread pudding protocols. In *Secure Information Networks*. Springer, 258–272.
[61] Benjamin Johnson, Aron Laszka, Jens Grossklags, Marie Vasek, and Tyler Moore. 2014. Game-theoretic analysis of DDoS attacks against Bitcoin mining pools. In *International Conference on Financial Cryptography and Data Security*. Springer, 72–86.
[62] Aljosha Judmayer, Nicholas Stifter, Philipp Schindler, and Edgar Weippl. 2018. Pitchforks in Cryptocurrencies: Enforcing rule changes through offensive forking-and. (2018).
[63] Aljosha Judmayer, Nicholas Stifter, Alexei Zamyatin, Itay Tsabary, Ittay Eyal, Peter Gaži, Sarah Meiklejohn, and Edgar Weippl. 2019. Pay-To-Win: Incentive Attacks on Proof-of-Work Cryptocurrencies. (2019).
[64] Dimitris Karakostas, Aggelos Kiayias, Christos Nasikas, and Dionysis Zindros. 2019. Cryptocurrency egalitarianism: a quantitative approach. *arXiv preprint arXiv:1907.02434* (2019).
[65] Aggelos Kiayias, Elias Koutsoupias, Maria Kyropoulou, and Yiannis Tselekounis. 2016. Blockchain mining games. In *Proceedings of the 2016 ACM Conference on Economics and Computation*. ACM, 365–382.
[66] Aggelos Kiayias, Alexander Russell, Bernardo David, and Roman Oliynykov. 2017. Ouroboros: A provably secure proof-of-stake blockchain protocol. In *Annual International Cryptology Conference*. Springer, 357–388.
[67] Joshua A Kroll, Ian C Davey, and Edward W Felten. 2013. The economics of Bitcoin mining, or Bitcoin in the presence of adversaries. In *Proceedings of WEIS*, Vol. 2013. 11.
[68] Yujin Kwon, Dohyun Kim, Yunmok Son, Eugene Vasserman, and Yongdae Kim. 2017. Be selfish and avoid dilemmas: Fork after withholding (faw) attacks on bitcoin. In *Proceedings of the 2017 ACM SIGSAC Conference on Computer and Communications Security*. ACM, 195–209.
[69] Yujin Kwon, Hyoungshick Kim, Jinwoo Shin, and Yongdae Kim. 2019. Bitcoin vs. Bitcoin Cash: Coexistence or Downfall of Bitcoin Cash? *arXiv preprint arXiv:1902.11064* (2019).
[70] Susan Landau. 2013. Making sense from Snowden: What's significant in the NSA surveillance revelations. *IEEE Security & Privacy* 11, 4 (2013), 54–63.
[71] Aron Laszka, Benjamin Johnson, and Jens Grossklags. 2015. When bitcoin mining pools run dry. In *International Conference on Financial Cryptography and Data Security*. Springer, 63–77.
[72] Xiaoqi Li, Peng Jiang, Ting Chen, Xiapu Luo, and Qiaoyan Wen. 2017. A survey on the security of blockchain systems. *Future Generation Computer Systems* (2017).
[73] Kevin Liao and Jonathan Katz. 2017. Incentivizing blockchain forks via whale transactions. In *International Conference on Financial Cryptography and Data Security*. Springer, 264–279.
[74] Loi Luu, Ratul Saha, Inian Parameshwaran, Prateek Saxena, and Aquinas Hobor. 2015. On power splitting games in distributed computation: The case of bitcoin pooled mining. In *2015 IEEE 28th Computer Security Foundations Symposium*. IEEE, 397–411.
[75] P. H. Madore. 2018. Bitcoin Cash: Craig Wright's BSV Suffers Multi-Block Reorg. https://www.ccn.com/competing-blockchains-bitcoin-cash-sv-reorganizes/
[76] Yuval Marcus, Ethan Heilman, and Sharon Goldberg. 2018. Low-Resource Eclipse Attacks on Ethereum's Peer-to-Peer Network. *IACR Cryptology ePrint Archive* 2018 (2018), 236.
[77] Francisco J. Marmolejo-Cossío, Eric Brigham, Benjamin Sela, and Jonathan Katz. 2019. Competing (Semi-)Selfish Miners in Bitcoin. In *Proceedings of the 1st ACM Conference on Advances in Financial Technologies* (Zurich, Switzerland) *(AFT '19)*. ACM, New York, NY, USA, 89–109. https://doi.org/10.1145/3318041.3355471
[78] Tim Maurer. 2018. *Cyber mercenaries*. Cambridge University Press.
[79] Patrick McCorry, Alexander Hicks, and Sarah Meiklejohn. 2018. Smart contracts for bribing miners. In *International Conference on Financial Cryptography and Data Security*. Springer, 3–18.
[80] James McWhinney. 2019. Why Governments Are Afraid of Bitcoin. https://www.investopedia.com/articles/forex/042015/why-governments-are-afraid-bitcoin.asp
[81] Dmitry Meshkov, Alexander Chepurnoy, and Marc Jansen. 2017. Short paper: Revisiting difficulty control for blockchain systems. In *Data Privacy Management, Cryptocurrencies and Blockchain Technology*. Springer, 429–436.
[82] Andrew Miller. 2013. Feather-forks: enforcing a blacklist with sub-50% hash power. https://bitcointalk.org/index.php?topic=312668.0
[83] Andrew Miller and Joseph J LaViola Jr. 2014. Anonymous byzantine consensus from moderately-hard puzzles: A model for bitcoin. *Available on line: http://nakamotoinstitute. org/research/anonymous-byzantine-consensus* (2014).
[84] ASIC miner value. 2019. Miners Profitability. https://www.asicminervalue.com/
[85] Bernhard Mueller. 2018. DoS with Block Gas Limit. https://github.com/ethereum/wiki/wiki/Safety#dos-with-block-gas-limit
[86] Bernhard Mueller. 2018. DoS with (Unexpected) Throw. https://github.com/ethereum/wiki/wiki/Safety#dos-with-unexpected-throw
[87] Phil Muncaster. 2017. World's Largest Bitcoin Exchange Bitfinex Crippled by DDoS. https://www.infosecurity-magazine.com/news/worlds-largest-bitcoin-exchange/
[88] Satoshi Nakamoto et al. 2008. Bitcoin: A peer-to-peer electronic cash system. (2008).
[89] Kartik Nayak, Srijan Kumar, Andrew Miller, and Elaine Shi. 2016. Stubborn mining: Generalizing selfish mining and combining with an eclipse attack. In *2016 IEEE European Symposium on Security and Privacy (EuroS&P)*. IEEE, 305–320.
[90] Bitcoin News. 2019. Bitmain Launches Low-Cost Special Edition Antminer S9. https://news.bitcoin.com/bitmain-launches-low-cost-special-edition-antminer-s9/
[91] Jianyu Niu and Chen Feng. 2019. Selfish Mining in Ethereum. *arXiv preprint arXiv:1901.04620* (2019).
[92] Shunya Noda, Kyohei Okumura, and Yoshinori Hashimoto. 2019. A Lucas Critique to the Difficulty Adjustment Algorithm of the Bitcoin System. *Available at SSRN 3410460* (2019).
[93] Stephen O'Neal. 2018. ABC vs SV: Assessing the Consequences of the Bitcoin Cash War. https://cointelegraph.com/news/abc-vs-cv-assessing-the-consequences-of-the-bitcoin-cash-war
[94] Jun Osawa. 2017. The escalation of state sponsored cyberattack and national cyber security affairs: is strategic cyber deterrence the key to solving the problem? *Asia-Pacific Review* 24, 2 (2017), 113–131.
[95] Rafael Pass, Lior Seeman, and Abhi Shelat. 2017. Analysis of the blockchain protocol in asynchronous networks. In *Annual International Conference on the Theory and Applications of Cryptographic Techniques*. Springer, 643–673.
[96] A Hash Pool. 2017. A Hash Pool. https://www.ahashpool.com/
[97] Bitcoin Project. 2015. Some Miners Generating Invalid Blocks. https://bitcoin.org/en/alert/2015-07-04-spv-mining
[98] Fabian Ritz and Alf Zugenmaier. 2018. The impact of uncle rewards on selfish mining in ethereum. In *2018 IEEE European Symposium on Security and Privacy Workshops (EuroS&PW)*. IEEE, 50–57.
[99] Eric Roberts. 1999. Multi-Person Prisoner's Dilemma. https://cs.stanford.edu/people/eroberts/courses/soco/projects/1998-99/game-theory/npd.html
[100] Meni Rosenfeld. 2011. Analysis of bitcoin pooled mining reward systems. *arXiv preprint arXiv:1112.4980* (2011).
[101] Meni Rosenfeld. 2014. Analysis of hashrate-based double spending. *arXiv preprint arXiv:1402.2009* (2014).
[102] Ayelet Sapirshtein, Yonatan Sompolinsky, and Aviv Zohar. 2016. Optimal selfish mining strategies in bitcoin. In *International Conference on Financial Cryptography and Data Security*. Springer, 515–532.
[103] SECBIT. 2018. How the winner got Fomo3D prize — A Detailed Explanation. https://medium.com/coinmonks/how-the-winner-got-fomo3d-prize-a-detailed-explanation-b30a69b7813f
[104] SFOX. 2019. Bitcoin Cash vs. Bitcoin SV: Six Months after the Hash War. https://blog.sfox.com/bitcoin-cash-vs-bitcoin-sv-six-months-after-the-hash-war-e6d92a03b891
[105] Scott Shane. 2017. The fake Americans Russia created to influence the election. *The New York Times* 7, 09 (2017).
[106] Atul Singh et al. 2006. Eclipse attacks on overlay networks: Threats and defenses. In *In IEEE INFOCOM*. Citeseer.
[107] Emil Sit and Robert Morris. 2002. Security considerations for peer-to-peer distributed hash tables. In *International Workshop on Peer-to-Peer Systems*. Springer, 261–269.
[108] Jakub Sliwinski and Roger Wattenhofer. [n.d.]. Blockchains Cannot Rely on Honesty. In *The 19th International Conference on Autonomous Agents and Multiagent Systems (AAMAS 2020)*.
[109] SmartMine. 2019. SmartMine – An intelligent way to mine cryptocurrency. https://www.smartmine.org/
[110] Joel Sobel and Ichiro Takahashi. 1983. A multistage model of bargaining. *The Review of Economic Studies* 50, 3 (1983), 411–426.
[111] Yonatan Sompolinsky and Aviv Zohar. 2015. Secure high-rate transaction processing in bitcoin. In *International Conference on Financial Cryptography and Data Security*. Springer, 507–527.
[112] Yonatan Sompolinsky and Aviv Zohar. 2018. Bitcoin's underlying incentives. *Commun. ACM* 61, 3 (2018), 46–53.
[113] Alexander Spiegelman, Idit Keidar, and Moshe Tennenholtz. 2018. Game of coins. *arXiv preprint arXiv:1805.08979* (2018).
[114] JOE STEWART. 2014. BGP Hijacking for Cryptocurrency Profit. https://www.secureworks.com/research/bgp-hijacking-for-cryptocurrency-profit
[115] Jason Teutsch, Sanjay Jain, and Prateek Saxena. 2016. When cryptocurrencies mine their own business. In *International Conference on Financial Cryptography and Data Security*. Springer, 499–514.
[116] Itay Tsabary and Ittay Eyal. 2018. The gap game. In *Proceedings of the 2018 ACM SIGSAC Conference on Computer and Communications Security*. ACM, 713–728.
[117] Itay Tsabary, Alexander Spiegelman, and Ittay Eyal. 2019. HEB: Hybrid Expenditure Blockchain. *arXiv* (2019), arXiv–1911.
[118] Marie Vasek, Micah Thornton, and Tyler Moore. 2014. Empirical analysis of denial-of-service attacks in the Bitcoin ecosystem. In *International conference on financial cryptography and data security*. Springer, 57–71.
[119] Bitcoin Wiki. 2017. Difficulty. https://en.bitcoin.it/wiki/Difficulty

[120] Shawn Wilkinson, Tome Boshevski, Josh Brandoff, and Vitalik Buterin. 2014. Storj a peer-to-peer cloud storage network. (2014).
[121] Shawn Wilkinson, Jim Lowry, and Tome Boshevski. 2014. Metadisk a blockchain-based decentralized file storage application. *Tech. Rep.* (2014).
[122] Fredrik Winzer, Benjamin Herd, and Sebastian Faust. 2019. Temporary censorship attacks in the presence of rational miners. In *2019 IEEE European Symposium on Security and Privacy Workshops (EuroS&PW)*. IEEE, 357–366.
[123] Gavin Wood et al. 2014. Ethereum: A secure decentralised generalised transaction ledger. *Ethereum project yellow paper* 151, 2014 (2014), 1–32.
[124] Shuangke Wu, Yanjiao Chen, Minghui Li, Xiangyang Luo, Zhe Liu, and Lan Liu. 2020. Survive and Thrive: A Stochastic Game for DDoS Attacks in Bitcoin Mining Pools. *IEEE/ACM Transactions on Networking* (2020).
[125] Fan Zhang, Ittay Eyal, Robert Escriva, Ari Juels, and Robbert Van Renesse. 2017. {REM}: Resource-Efficient Mining for Blockchains. In *26th {USENIX} Security Symposium ({USENIX} Security 17)*. 1427–1444.
[126] Wolfie Zhao. 2019. Bitcoin Miners Halt Operations as Rainstorm Triggers Mudslides in China. https://www.coindesk.com/bitcoin-miners-halt-operations-as-rainstorm-triggers-fatal-mudslide-in-china
[127] Wolfie Zhao. 2020. Older Mining Machines Turn Profitable Again as Bitcoin Rises Ahead of Halving. https://www.coindesk.com/older-mining-machines-turn-profitable-again-as-bitcoin-rises-ahead-of-halving

## A CHANGING ACTION IN THE MIDDLE OF THE ROUND

In the model, we assumed that no rational miner changes her action in the middle of the round. We now justify this assumption. As mentioned earlier, the coin price is assumed to be constant during the entire game. Therefore, the honest game profitability factor $\omega_i^b$ of $\mathcal{P}_i$ keeps its value constant during the round. In addition, we assume that no miner withholds blocks. We define as $\text{Time}_j$ the time when round $j$ ends and round $j+1$ starts.

**Claim A.1.** *If $\mathcal{P}_i$ chooses an action $a$ in the beginning of round $j$ ($\text{Time}_{j-1}$), she does not gain anything from changing her action for all $t$ that hold $\text{Time}_{j-1} < t < \text{Time}_j$.*

Proof. We know that the rational miner $\mathcal{P}_i$ chose the most beneficial action $a$ in the beginning of round $r$, assume by contradiction that it is beneficial for $\mathcal{P}_i$ to change her action in time $t_1$ that holds $\text{Time}_{j-1} < t_1 < \text{Time}_j$ to a different action $a'$ s.t $a \neq a'$. Previous works showed that new block appearance in the system can be described with Poisson distribution, with the time between blocks correspond to exponential distribution [111]. One of the properties of this distribution is that it is memoryless. Since $\mathcal{P}_i$ has the same probability of finding a new block as she had at the beginning of the round (and so do other miners), she has the same expected revenue from each action. If changing action in the middle of a round is profitable, this implies that changing an action was also beneficial at the beginning of the round. This is a contradiction to the fact that $\mathcal{P}_i$ is rational and chose the best action at the beginning of the round. □

Note that for memorylessness, we had to assume that there is no block withholding in the system, i.e., in every point during the round, it is known by everyone that there was no new block mined, by any miner, since the beginning of the round. For example, this assumption does not hold when there is an active selfish mining attack [46]. Although, it is reasonable to assume that no miner is withholding blocks during the attack as there is no evidence of cases of selfish mining attacks in the wild.

## B MINE IN STATE 0 AND STATE 2 ALWAYS BETTER ACTION THAN STOP

Throughout the paper, we assume that miners always play mine in State 0 and State 2. We now prove formally that mine is always better action than stop in these states. In other words, assume that there are two strategies that differ only by the action in state 0 (or state 2), namely strategy $S_A$ uses action mine while strategy $S_B$ uses stop. It necessary means that $U_{S_A} > U_{S_B}$.

**Claim B.1.** *If $\omega_i^b > 1$ then mine in state 0 and state 2 is always more profitable than stop for $\mathcal{P}_i$.*

Proof. We show the claim for state 0. The proof for state 2 is the equivalent. As we did before we consider two strategies $S_A$ and $S_B$ that differ only by the action in state 0 (mine for $S_A$ vs. stop for $S_B$). We need to compare the utilities of two strategies that differ only by the action of $\mathcal{P}_i$ in state 0. First, we observe that $p_1$ does not change as a result of $\mathcal{P}_i$'s action in state 0. This is because the rate from state 0 to state 1 and the rate from state 1 to state 2 are not affected by wether or not $\mathcal{P}_i$ mines in state 0 (or state 2). We denote by $\rho_s$ the normalized profit rate in state $s$, it is equal to the product of the expected block reward, and the normalized rate $\mathcal{P}_i$ finds blocks. We denote by $\rho_1$ and $\rho_2$ the expected profit rates in states 1 and 2 respectively. With $\rho_{\text{mine}}$ and $\rho_{\text{stop}}$ the profit rates of playing mine and stop in state 0 respectively. We denote by $p_0$ and $p_2$ the state probabilities of state 0 and state 2, respectively, when playing mine in state 0. We denote by $p_0'$ and $p_2'$ the state probabilities of state 0 and state 2 respectively when playing stop in state 0. Therefore the utility of playing mine in state 0 is:

$$U_{\text{mine}} = \rho_{\text{mine}} \cdot p_0 + \rho_1 \cdot p_1 + \rho_2 \cdot p_2.$$

Similarly, the utility of playing stop in state 0 is:

$$U_{\text{stop}} = \rho_{\text{stop}} \cdot p_0' + \rho_1 \cdot p_1 + \rho_2 \cdot p_2'.$$

The profit rates in state 1 and state 2 ($\rho_1$ and $\rho_2$) can not be larger than the profit rate in state 0 ($\rho_{\text{mine}}$) as $\rho_{\text{mine}}$ is the maximal possible profit rate. Therefore, it holds that $\rho_{\text{mine}} \geq \rho_1, \rho_2$ and $\rho_{\text{mine}} > 0$ (as $\omega_i^b > 1$). Additionally, there is no reward and cost when not mining, so $\rho_{\text{stop}} = 0$. Thus, the following inequality holds:

$$U_{\text{mine}} = \rho_{\text{mine}} \cdot p_0 + \rho_1 \cdot p_1 + \rho_2 \cdot p_2 \geq \rho_1 \cdot p_1 + \rho_2 \cdot (p_0 + p_2)$$
$$= \rho_1 \cdot p_1 + \rho_2 \cdot (p_0' + p_2') > \rho_1 \cdot p_1 + \rho_2 \cdot p_2' = U_{\text{stop}}.$$
□

## C COST OF 51% ATTACK

We show our calculation for the cost of 51% attack. In March 2020, the total hash rate of Bitcoin is roughly 120,000,000 TH/s. The most advanced mining equipment is considered to be Bitmain S17 Pro which has hashrate of 50 TH/s and power consumption of 1.975 kW [15]. The official cost of a unit is $2128. Another widely used ASIC machine, which is significantly cheaper to acquire, is Bitmain S9 SE [90]. The hash rate of this machine is 16 TH/s; its power consumption is 1.280 h and unit price $350. The number of S17 Pro rigs required to have the majority of mining power in the network is: $\lceil \frac{120,000,000}{50} \rceil = 2,400,000$. With total cost of $2,400,000 \cdot 2128 = \$5B$ and power consumption of $2,400,000 \cdot 1.975 = 4,740,000$ kW which with electricity price of $0.04/kWh and additional 15% overhead

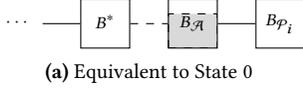

(a) Equivalent to State 0

Figure 9: *States*

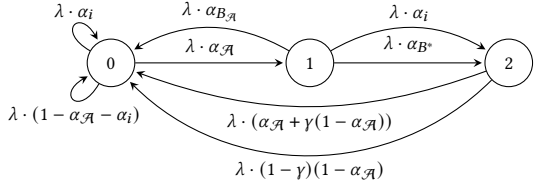

(a) $S_{\text{mine}}$: $\mathcal{P}_i$ mines on $B^*$ in state 1

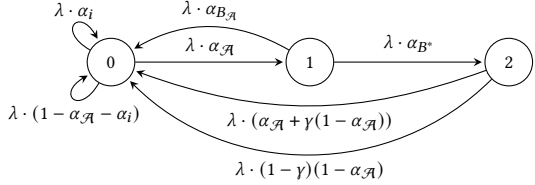

(b) $S_{\text{stop}}$: $\mathcal{P}_i$ stops mining in state 1

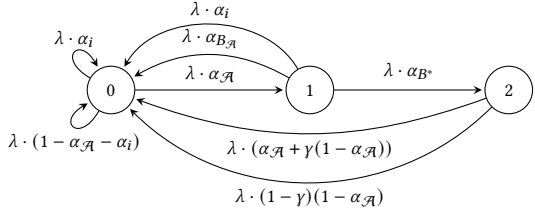

(c) $S_{\text{SPV}}$: $\mathcal{P}_i$ mines on $B_{\mathcal{A}}$ in state 1

Figure 10: *Markov chains*

OPEX expenses would cost \$5.2M a day. Similarly, for S9 SE, the equipment cost would be \$2.7B, and the daily electricity cost would be \$10.6M.

## D BDOS ATTACK WITH SPV MINING

In this appendix we give the full formal analysis of BDoS in the presence of SPV mining, that was defined in §6.

### D.1 Analysis Changes

As before we denote by $\Lambda_{B_{\mathcal{A}}}$ the set of miners actively mining on $B_{\mathcal{A}}$ while the attack is active. We define: $\alpha_{B_{\mathcal{A}}} \triangleq \sum_{j \in \Lambda_{B_{\mathcal{A}}}} \alpha_j$. Next, in the same way as in eq. (3) we denote by $\alpha_{B_{\mathcal{A}}}(S)$ the mining power of miners that mine on $B_{\mathcal{A}}$ in state 1:

$$\alpha_{B_{\mathcal{A}}}(S) \triangleq \begin{cases} \alpha_{B_{\mathcal{A}}} + \alpha_i, & \text{if } S = S_{\text{SPV}} \\ \alpha_{B_{\mathcal{A}}}, & \text{otherwise.} \end{cases}$$

To analyze the dominant strategy, we construct Markov chains for each strategy in the new strategy space $\{S_{\text{stop}}, S_{\text{mine}}, S_{\text{SPV}}\}$, similarly to what was done in §5. The Markov Chains (fig. 10) for $S_{\text{stop}}$ and $S_{\text{mine}}$ are similar to the ones shown in fig. 2. Although, the chains described in fig. 10 have an extra edge from state 1 to state 0 with rate $\lambda \cdot \alpha_{B_{\mathcal{A}}}$ that corresponds to the portion of rational miners (excluding $\mathcal{P}_i$) that keep mining during the attack. In fig. 10c there is an additional edge from state 1 to state 0 with rate $\lambda \cdot \alpha_i$, as now $\mathcal{P}_i$ also mines during the attack.

We now can calculate the state probabilities for each strategy. We denote the states probabilities with $p_0^S$, $p_1^S$ and $p_2^S$ for states 0, 1 and 2 respectively (as in §5). Recall that states 0, 1 and 2 correspond to the initial, attack is progress and race states respectively. We denote the utility functions by $U_i^{S_{\text{stop}}}$, $U_i^{S_{\text{mine}}}$ and $U_i^{S_{\text{SPV}}}$ for $S_{\text{stop}}$, $S_{\text{mine}}$ and $S_{\text{SPV}}$ respectively. The Markov chains for $S_{\text{mine}}$ and $S_{\text{stop}}$ are almost identical to the ones described in fig. 2, with a new edge from state 1 to state 0 that corresponds to a portion $\alpha_{B_{\mathcal{A}}}$ of other miners that mine on $\mathcal{A}$'s block. The Markov chain for $S_{\text{SPV}}$ is similar to the new Markov chain for $S_{\text{stop}}$ but with an edge from state 1 to state 0 that corresponds to $\mathcal{P}_i$'s efforts to extend $B_{\mathcal{A}}$.

Next we calculate the state probabilities for each state depending on the strategy using basic Markov chains analytical analysis:

$$\begin{aligned} p_0^S &= \frac{\alpha_{B^*}(S) + \alpha_{B_{\mathcal{A}}}(S)}{\alpha_{\mathcal{A}} \cdot \alpha_{B^*}(S) + \alpha_{\mathcal{A}} + \alpha_{B_{\mathcal{A}}}(S) + \alpha_{B^*}(S)}, \\ p_1^S &= \frac{\alpha_{\mathcal{A}}}{\alpha_{\mathcal{A}} \cdot \alpha_{B^*}(S) + \alpha_{\mathcal{A}} + \alpha_{B_{\mathcal{A}}}(S) + \alpha_{B^*}(S)}, \\ p_2^S &= \frac{\alpha_{\mathcal{A}} \cdot \alpha_{B^*}(S)}{\alpha_{\mathcal{A}} \cdot \alpha_{B^*}(S) + \alpha_{\mathcal{A}} + \alpha_{B_{\mathcal{A}}}(S) + \alpha_{B^*}(S)}. \end{aligned} \quad (10)$$

The utilities for $S_{\text{stop}}$ and $S_{\text{mine}}$ are identical to the ones in 5 and 6 in respect to state probabilities, as non of the original edges, where $\mathcal{P}_i$ gets a reward, have changed:

$$\begin{aligned} U_i^{S_{\text{stop}}} &= (p_0^{S_{\text{stop}}} + p_2^{S_{\text{stop}}}) \cdot \lambda K - (1 - p_1^{S_{\text{stop}}}) \cdot c_i, \\ U_i^{S_{\text{mine}}} &= (p_0^{S_{\text{mine}}} + p_2^{S_{\text{mine}}} + (1-\gamma)(1-\alpha_{\mathcal{A}}) \cdot p_1^{S_{\text{mine}}})\lambda K - c_i. \end{aligned}$$

Finally, we calculate the utility for playing $S_{\text{SPV}}$:

$$\begin{aligned} U_i^{S_{\text{SPV}}} &= \frac{1}{\alpha_i}(\hat{R}_i^{S_{\text{SPV}}} - \hat{C}_i^{S_{\text{SPV}}}) \\ &= (p_0^{S_{\text{SPV}}} + p_2^{S_{\text{SPV}}}) \cdot \lambda K - c_i. \end{aligned} \quad (11)$$

### D.2 Narrowing down the possible number of strategies

In order to simplify the analysis, we spot a dominated strategy, i.e., a strategy that is always less beneficial compared to another strategy.

**Claim D.1.** *$S_{SPV}$ is strictly dominated by $S_{mine}$.*

Proof. We calculate the difference $\Delta$ between the utility of playing $S_{\text{mine}}$ (defined in eq. (6)) and the utility of playing $S_{\text{SPV}}$ (defined in eq. (11)):

$$\begin{aligned} \Delta &\triangleq U_i^{S_{\text{mine}}} - U_i^{S_{\text{SPV}}} \\ &= p_0^{S_{\text{mine}}} + p_2^{S_{\text{mine}}} + (1-\gamma)(1-\alpha_{\mathcal{A}}) \cdot p_1^{S_{\text{mine}}} \\ &\quad - (p_0^{S_{\text{SPV}}} + p_2^{S_{\text{SPV}}})) \cdot \lambda K. \end{aligned} \quad (12)$$

We note that the probability $p_1^S$ (eq. (10)) decreases when $\mathcal{P}_i$ chooses $S_{\text{mine}}$ instead of $S_{\text{SPV}}$, the numerator stays the same while the denominator increases. We conclude that $p_1^{S_{\text{mine}}} < p_1^{S_{\text{SPV}}}$ and therefore:

$$(p_0^{S_{\text{mine}}} + p_2^{S_{\text{mine}}}) - (p_0^{S_{\text{SPV}}} + p_2^{S_{\text{SPV}}}) \quad (13)$$
$$= (1 - p_1^{S_{\text{mine}}}) - (1 - p_1^{S_{\text{SPV}}}) = p_1^{S_{\text{SPV}}} - p_1^{S_{\text{mine}}} > 0.$$

From eq. (12) and eq. (13) we conclude that $\Delta > 0$. Therefore by playing $S_{\text{mine}}$, $\mathcal{P}_i$ always has a strictly larger profit than she would have if she would play $S_{\text{SPV}}$. □

From now on we consider only two strategies for $\mathcal{P}_i$ in our analysis: $S_{\text{mine}}$ and $S_{\text{stop}}$, as we proved that $\mathcal{P}_i$ never chooses strategy $S_{\text{SPV}}$. Note that we still have to consider $S_{\text{SPV}}$ for other miners in order to find conditions for $S_{\text{stop}}$ to be dominant strategy (section D.3). In section D.4 we relax this in order to argue about the more practical setting where no rational miner chooses a dominated strategy.

### D.3 Conditions for Successful Attack

As in section 5.1 we calculate for what values of $\omega_i^b$ the attack would be successful given $\alpha_{\mathcal{A}}$ and $\alpha_i$. We define $D(\alpha_{B^*}, \alpha_{B_{\mathcal{A}}})$ to be the normalized difference between $U_i^{S_{\text{stop}}}$ and $U_i^{S_{\text{mine}}}$:

$$D(\alpha_{B^*}, \alpha_{B_{\mathcal{A}}}) \triangleq \frac{U_i^{S_{\text{stop}}} - U_i^{S_{\text{mine}}}}{c_i}$$
$$= (p_0^{S_{\text{stop}}} + p_2^{S_{\text{stop}}} - p_0^{S_{\text{mine}}} - p_2^{S_{\text{mine}}} \quad (14)$$
$$- (1 - \gamma)(1 - \alpha_{\mathcal{A}}) \cdot p_1^{S_{\text{mine}}}) \cdot \omega_i^b + p_1^{S_{\text{stop}}}.$$

As before we find for values of $\omega_i^b$ for all possible $\alpha_{B^*}$ and $\alpha_{B_{\mathcal{A}}}$ it holds that $D(\alpha_{B^*}, \alpha_{B_{\mathcal{A}}}) > 0$. We therefore calculate the condition on $\omega_i^b$ so that $D(\alpha_{B^*}, \alpha_{B_{\mathcal{A}}}) > 0$ using eq. (14):

$$\omega_i^b < \underbrace{\frac{p_1^{S_{\text{stop}}}}{p_0^{S_{\text{mine}}} + p_2^{S_{\text{mine}}} + (1 - \gamma)(1 - \alpha_{\mathcal{A}}) \cdot p_1^{S_{\text{mine}}} - (p_0^{S_{\text{stop}}} + p_2^{S_{\text{stop}}})}}_{Q(\alpha_{B^*}, \alpha_{B_{\mathcal{A}}})}.$$
$$(15)$$

This is the general bound on $\omega_i^b$ that makes $S_{\text{stop}}$ the dominant strategy for $\mathcal{P}_i$. This can be solved for specific values of $\gamma$, $\alpha_{\mathcal{A}}$ and $\alpha_i$ and otherwise it's not analytically solvable for the parametric case.

### D.4 Iterated Elimination of Strictly Dominated Strategies

The result in eq. (15) is the condition for $S_{\text{stop}}$ to be strictly dominating strategy among the three strategies: $\{S_{\text{stop}}, S_{\text{mine}}, S_{\text{SPV}}\}$. We use a technique called iterated elimination of strictly dominated strategies (IESDS) [47] and show that our game is *dominance-solvable game*. We assume that no rational miner chooses to mine on $B_{\mathcal{A}}$ and that this is a common knowledge that no other miner would mine on it [6], as this is a strictly dominated strategy as we showed in section D.2. This elimination would leave us with the only Nash equilibrium in the game. Therefore, we analyze the case where $\alpha_{B_{\mathcal{A}}} = 0$. This implies that if the result in eq. (9) holds for all rational miners, $S_{\text{stop}}$ is the *only Nash equilibrium* in the game [67]. This equilibrium is conceptually stronger than general equilibrium, as it implies that $S_{\text{stop}}$ is the best strategy regardless of other miners' *rational* strategies.

---

**Algorithm 1** Scheduler

1: $r \leftarrow 0$
2: 
3: **loop** // The scheduler runs in an infinite loop.
4:    $r \leftarrow r + 1$
5:    active $\leftarrow \emptyset$
6:    **for** $p \in \{\mathcal{A}, \mathcal{P}_1, \ldots, \mathcal{P}_n\}$ **do**
7:      **if** $p$.Mine_This_Round = true **then**
8:        template$_p \leftarrow p$.Get_Block_Template
9:        active $\leftarrow$ active $\bigcup p$
10:      **end if**
11:    **end for**
12:    $T \leftarrow$ Exp_Distribution($\lambda \cdot \sum_{p \in \text{active}} \alpha_p$)
13:    sleep($T$) // Simulate block time.
14:    w $\leftarrow$ Sample by weight of hashrate from active
15:    $B_r \leftarrow$ Generate_Valid_Block($r$, template$_w$)
16: 
17:    **if** w = $\mathcal{A}$ **then**
18:      publish $\leftarrow \mathcal{A}$.Find_New_Block($B_r$)
19:      **if** publish = "header" **then**
20:        $H$ = Get_Header($B_r$)
21:        **for** $p \in \{\mathcal{P}_1, \ldots, \mathcal{P}_n\}$ **do** $p$.Add_Header($H$)
22:      **else if** publish = "full block" **then**
23:        **for** $p \in \{\mathcal{A}, \mathcal{P}_1, \ldots, \mathcal{P}_n\}$ **do** $p$.Add_Block($B_r$)
24:      **end if**
25:    **else**
26:      competing $\leftarrow \mathcal{A}$.Get_Competing_Blocks($B_r$)
27:      **if** competing.empty = true **then**
28:        **for** $p \in \{\mathcal{A}, \mathcal{P}_1, \ldots, \mathcal{P}_n\}$ **do** $p$.Add_Block($B_r$)
29:      **else**
30:        Send_Blocks(w, $[B_r]$ + competing)
31:        Send_Blocks($\mathcal{A}$, competing + $[B_r]$)
32:        **for** $p \in \{\mathcal{P}_1, \ldots, \mathcal{P}_n\} \setminus \{w\}$ **do**
33:          **with** probability $\frac{\gamma(1-\alpha_{\mathcal{A}})}{1-\alpha_{\mathcal{A}}-\alpha_w}$:
34:            Send_Blocks($p$, competing + $[B_r]$)
35:          **with** probability $1 - \frac{\gamma(1-\alpha_{\mathcal{A}})}{1-\alpha_{\mathcal{A}}-\alpha_w}$:
36:            Send_Blocks($p$, $[B_r]$ + competing)
37:        **end for**
38:      **end if**
39:    **end if**
40: **end loop**
41: 
42: **function** Send_Blocks($p$, blocks)
43:    **for** $B \in$ blocks **do** $p$.Add_Block($B$)
44: **end function**

---

## E PSEUDO-CODE FOR MODEL

In this section, we describe the pseudo-code for the scheduler (algorithm 1), adversary (algorithm 2) and the rational miner (algorithm 3) that were described in section 3.1. Note that for simplicity of the pseudocode we denote the mining power of rational miner $\mathcal{P}_i$ as $\alpha_{\mathcal{P}_i}$ as well, so $\alpha_{\mathcal{P}_i} := \alpha_i$.

### Algorithm 2 Adversary $\mathcal{A}$

1: $L_{\mathcal{A}} \leftarrow \{B_0\}, O_{\mathcal{A}}[B_0] \leftarrow 0, r \leftarrow 0$
2: $B_{\text{withheld}} \leftarrow \bot, B_{\text{extend}} \leftarrow B_0$
3:
4: **function** Mine_This_Round
5:    $r \leftarrow r + 1$
6:    **if** $B_{\text{withheld}} = \bot$ **then**
7:       **return** true
8:    **else**
9:       **return** false
10:    **end if**
11: **end function**
12:
13: **function** Get_Block_Template
14:    **return** Generate_Template($\mathcal{A}$, Get_Header($B_{\text{extend}}$))
15: **end function**
16:
17: **function** Find_New_Block($B$)
18:    $B_{\text{withheld}} \leftarrow B$
19:    **return** "header"
20: **end function**
21:
22: **function** Get_Competing_Blocks($B$)
23:    **if** Get_Height($B$) = Get_Height($B_{\text{withheld}}$) **then**
24:       $B_{\text{withheld}} \leftarrow \bot$
25:       **return** $[B_{\text{withheld}}]$
26:    **else**
27:       **return** [ ]
28:    **end if**
29: **end function**
30:
31: **function** Add_Block($B$)
32:    $L_{\mathcal{A}} \leftarrow L_{\mathcal{A}} \cup \{B\}, O_{\mathcal{A}}[B] \leftarrow |L_{\mathcal{A}}|$
33:    **if** $B = B_{\text{withheld}}$ **then**
34:       $B_{\text{withheld}} \leftarrow \bot$
35:    **end if**
36:    **if** Get_Height($B$) > Get_Height($B_{\text{extend}}$) **then**
37:       $B_{\text{extend}} \leftarrow B$
38:    **end if**
39: **end function**

### Algorithm 3 Rational Player $\mathcal{P}_i$

1: $L_i \leftarrow \{B_0\}, O_i[B_0] \leftarrow 0, r \leftarrow 0$
2: $B_{\text{header}} \leftarrow \bot, B_{\text{extend}} \leftarrow B_0$
3: $M \leftarrow$ Get_Best_Strategy($BDoS, \alpha_i, \omega_i^b$)
4:
5: **function** Mine_This_Round
6:    $r \leftarrow r + 1$
7:    **if** $M[L_i][O_i]$ = stop **then**
8:       **return** false
9:    **else**
10:       **return** true
11:    **end if**
12: **end function**
13:
14: **function** Get_Block_Template
15:    **if** $M[L_i][O_i]$ = mineSPV **then**
16:       **return** Generate_Template($\mathcal{P}_i$, Get_Header($B_{\text{header}}$))
17:    **else if** $M[L_i][O_i]$ = mine **then**
18:       **return** Generate_Template($\mathcal{P}_i$, Get_Header($B_{\text{extend}}$))
19:    **end if**
20: **end function**
21:
22: **function** Add_Block($B$)
23:    $L_{\mathcal{A}} \leftarrow L_{\mathcal{A}} \cup \{B\}, O_{\mathcal{A}}[B] \leftarrow |L_{\mathcal{A}}|$
24:    **if** Get_Header($B$) = Get_Header($B_{\text{header}}$) **then**
25:       $B_{\text{header}} \leftarrow \bot$
26:    **end if**
27:    **if** Get_Height($B$) > Get_Height($B_{\text{extend}}$) **then**
28:       $B_{\text{extend}} \leftarrow B$
29:    **end if**
30: **end function**
31:
32: **function** Add_Header($H$)
33:    $B_{\text{header}} \leftarrow (H, \bot)$
34: **end function**

# F ANALYSIS WITH DIFFERENT ELECTRICITY PRICES

BDoS is sensitive to miners' profitability. In §8, we made a conservative assumption that the electricity price is $0.04/kWh: the higher the electricity price is, the easier it becomes for a BDoS attacker. Other sources [12, 39, 43, 92, 127] assume that $0.05/kWh or $0.06/kWh are the average prices miners pay for electricity.

In this section, we compute Bitcoin mining profitability and attacker's resource threshold for different electricity prices during the past year, assuming no altruistic miners ($\eta = 0$). We plot the result in fig. 11. As before, we are interested in finding the minimum threshold for the attack. This is due to the attacker's ability to choose the timing of the attack. We can see that the threshold for the attacker's hash rate was as low as 7.7% before halving and 10% after halving for $0.05/kWh and less than 1% for both before and after halving for $0.06/kWh.

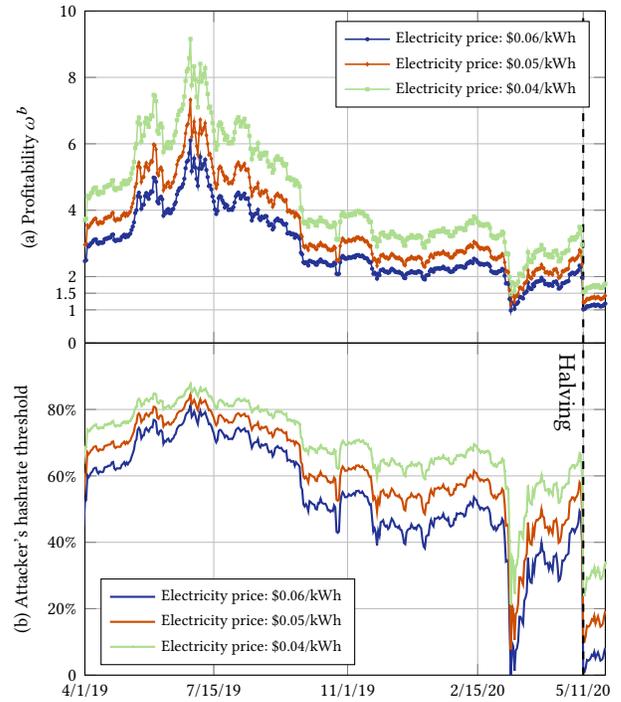

**Figure 11:** *(a) Profitability of mining Bitcoin; (b) Attacker's resource threshold.*